\documentclass[aps,pra,10pt,superscriptaddress,onecolumn,showkeys]{revtex4-2}
\pdfoutput=1
\usepackage[english]{babel}
\usepackage[utf8]{inputenc}
\usepackage[dvipsnames]{xcolor}
\usepackage{graphicx}
\usepackage{url}
\usepackage[colorlinks=true, urlcolor=Blue, citecolor=Red, linkcolor=Red, anchorcolor=Red]{hyperref}
\usepackage{bm}
\usepackage[labelfont=bf,justification=raggedright]{caption}
\usepackage{subcaption}
\usepackage{adjustbox}
\usepackage{mathtools}
\usepackage{amssymb}
\usepackage{amsfonts}
\usepackage{amsthm}
\usepackage{physics}
\usepackage{makecell}
\usepackage{dsfont}
\usepackage{amsmath}

\DeclareMathOperator*{\argmin}{arg\,min}
\DeclareMathOperator{\logit}{logit}
\newtheorem{theorem}{Theorem}
\newtheorem{innercustomtheorem}{Theorem}
\newenvironment{customtheorem}[1]
  {\renewcommand\theinnercustomtheorem{#1}\innercustomtheorem}
  {\endinnercustomtheorem}
\renewcommand{\epsilon}{\varepsilon}

\usepackage[ruled]{algorithm2e}
\begin{document}

\title{F-Divergences and Cost Function Locality in Generative Modelling with\\ Quantum Circuits}

\author{Chiara Leadbeater}
\thanks{These authors contributed equally to this work.}
\affiliation{Cambridge Quantum Computing Limited, SW1E 6DR London, United Kingdom}

\author{Louis Sharrock}
\thanks{These authors contributed equally to this work.}
\affiliation{Cambridge Quantum Computing Limited, SW1E 6DR London, United Kingdom}
\affiliation{Department of Mathematics, Imperial College London, London SW7 2AZ, United Kingdom}

\author{Brian Coyle}
\affiliation{Cambridge Quantum Computing Limited, SW1E 6DR London, United Kingdom}
\affiliation{School of Informatics, University of Edinburgh,  EH8 9AB Edinburgh, United Kingdom}

\author{Marcello Benedetti}
\email{marcello.benedetti@cambridgequantum.com}
\affiliation{Cambridge Quantum Computing Limited, SW1E 6DR London, United Kingdom}

\date{October 8, 2021}

\begin{abstract} 
Generative modelling is an important unsupervised task in machine learning. In this work, we study a hybrid quantum-classical approach to this task, based on the use of a quantum circuit Born machine. In particular, we consider training a quantum circuit Born machine using $f$-divergences. We first discuss the adversarial framework for generative modelling, which enables the estimation of any $f$-divergence in the near term. Based on this capability, we introduce two heuristics which demonstrably improve the training of the Born machine. The first is based on $f$-divergence switching during training. The second introduces locality to the divergence, a strategy which has proved important in similar applications in terms of mitigating barren plateaus. Finally, we discuss the long-term implications of quantum devices for computing $f$-divergences, including algorithms which provide quadratic speedups to their estimation. In particular, we generalise existing algorithms for estimating the Kullback-Leibler divergence and the total variation distance to obtain a fault-tolerant quantum algorithm for estimating another $f$-divergence, namely, the Pearson divergence.
\end{abstract}


\maketitle

\vspace{-5mm}

\section{Introduction}

One of the most challenging technological questions of our time is whether existing quantum computers can achieve quantum advantage in tasks of practical interest. Variational quantum algorithms (VQAs), which are well suited to the constraints imposed by existing devices, have emerged as the leading strategy for achieving such a quantum advantage~\cite{McClean2016,Benedetti2019,Cerezo2020a,Bharti2021}. 

In VQAs, a problem-specific cost function, which typically consists of a functional of the output of a parameterised quantum circuit, is efficiently evaluated using a quantum computer. Meanwhile, a classical optimiser is leveraged to train the circuit parameters in order to minimise the cost function. This hybrid quantum-classical approach is robust to the limited connectivity and qubit count of existing devices, and, by restricting the circuit depth, also provides an effective strategy for error mitigation. 

Given their flexibility, VQAs have been proposed for a vast array of applications. Of~particular relevance are applications of VQAs to machine learning problems, including classification~\cite{Li2021Review,Grant2018,Cong2019,Schuld2019,Havlicek2019,LaRose2020}, data compression~\cite{Romero2017,Pepper2019,Ding2019}, clustering~\cite{Otterbach2017}, generative modelling~\mbox{\cite{Liu2018,Benedetti2019a,Hamilton2019,Zhu2019,Coyle2020,Du2020,Anand2021a,Leyton-Ortega2021,Demers2018,Hu2019,Zeng2019,Zoufal2019,verdon2019quantum,Huang2020,Situ2020,Coyle2021,Liu2021,rudolph2020generation} }, and inference~\cite{Benedetti2021}.

In this paper, we focus on a hybrid quantum-classical approach to generative modelling using a Born machine~\cite{cheng2018information}. We adopt an adversarial framework to this task, in which a Born machine (the `generator') generates samples from the target distribution, while a binary classifier (the `discriminator') attempts to distinguish between generated samples and true samples. This is sometimes referred to in the literature as a quantum generative adversarial network.

In a generalisation of existing approaches, we consider training the Born machine with respect to any $f$-divergence as a cost function. Well-known examples of $f$-divergences include the Kullback-Leibler divergence (KL), the Jensen-Shannon divergence (JS), the squared Hellinger distance ($\text{H}^2$), the total variation distance (TV), and the Pearson divergence ($\chi^2$). In the adversarial framework, it is straightforward to estimate the $f$-divergence: any such divergence is defined in terms of the density ratio of the target distribution and model distribution, which can be estimated using standard techniques via the output of the binary classifier~\cite{Sugiyama2012}. On this basis, we propose a heuristic for training the Born machine, based on the idea of dynamically switching the $f$-divergence during training in order to optimise the rate of convergence and utilise favourable qualities of each one. We also propose a second heuristic, based on introducing locality into the $f$-divergence, motivated by the now well-established connection between locality and barren plateaus in VQA training landscapes~\cite{Cerezo2021,Uvarov2021}. For both heuristics, we provide numerical evidence to suggest that they can lead to (sometimes significant) performance improvements, particularly in under- and over-parameterised circuits.

We conclude this paper with a discussion of the longer-term implication of quantum devices for computing the $f$-divergences between two probability distributions. In particular, we discuss the existence of quadratic speedups for the estimation of TV and KL shown by~\cite{bravyi_quantum_2011, montanaro_quantum_2015, li_quantum_2019} and extend these results to an algorithm for estimating $\chi^2$, assuming access to a fault-tolerant quantum computer.

The remainder of this paper is organised as follows. In Section \ref{sec:backgroun}, we begin by introducing generative modelling, Born machines, and $f$-divergences. In Section \ref{sec:training_strategies}, we then introduce the two training heuristics for the Born machine. In Section \ref{sec:results}, we provide numerical results to demonstrate the performance of the heuristics. In Section \ref{sec:fault_tolerant_est}, we discuss the long-term implications of quantum devices for computing $f$-divergences. Finally, in Section \ref{sec:discussion}, we offer some concluding remarks.

\section{Background} 
\label{sec:backgroun}

\subsection{Generative Modelling}
\label{ssec:generative_modelling}

Generative modelling is an unsupervised machine learning task in which the goal is to learn the probability distribution which generates a given data set. More precisely, given access to i.i.d. samples $\smash{\boldsymbol{x}_1,\dots,\boldsymbol{x}_m\stackrel{\text{i.i.d.}}{\sim} p(\boldsymbol{x})}$ in $\mathbb{R}^p$, the objective of generative modelling is to learn a model $q_{\boldsymbol{\theta}}(\boldsymbol{x})$, typically parameterised by a $d$ dimensional parameter vector,  $\boldsymbol{\theta}\in\mathbb{R}^{d}$, which closely resembles $p(\boldsymbol{x})$. Generative models find applications in a wide range of problems, ranging from the typical modalities of machine learning such as text~\cite{Bowman2016}, image~\cite{Zhu2017} and graph~\cite{Simonovsky2018} analysis, to problems in active learning~\cite{Sinha2019}, reinforcement learning~\cite{Ha2018}, medical imaging~\cite{Ilse2020}, physics~\cite{Brehmer2020}, and speech synthesis~\cite{Oord2016}.

Broadly speaking, one can distinguish between two main categories of generative model: prescribed models and implicit models~\cite{Diggle1984,Mohamed2017}. Prescribed models provide an explicit parametric specification of the distribution of the observed random variable $\boldsymbol{x}$, directly specifying the density $q_{\boldsymbol{\theta}}(\boldsymbol{x})$. An example of a prescribed model is the ubiquitous multivariate Gaussian distribution. Implicit models, on the other hand, specify only the stochastic procedure which generates samples. An example of an implicit model is a complex computer simulation of some physical phenomenon, for which the likelihood function cannot be computed. Since, in this case, one no longer models $q_{\boldsymbol{\theta}}(\boldsymbol{x})$ directly, valid objectives can now only involve quantities (e.g., expectation values) which can be estimated efficiently using samples.

In the last three decades, a number of generative models, both explicit and implicit, have been proposed in the machine learning literature. These include autoregressive models~\cite{Frey1998,uria2016neural}, normalising flows~\cite{Rippel2013,Rezende2015,Dinh2017}, variational autoencoders~\cite{Kingma2014,Rezende2014}, Boltzmann machines~\cite{Ackley1985,Hinton2006,Salakhutdinov2009}, generative stochastic networks~\cite{Bengio2014}, generative moment matching networks~\cite{Dziugaite2015,Li2015a}, and generative adversarial networks~\cite{Goodfellow2014}. These models are classically implemented using deep neural network architectures. In recent years, however, hybrid quantum-classical approaches based on parameterised quantum circuits have also gained traction~\cite{Liu2018,Benedetti2019a,Hamilton2019,Zhu2019,Coyle2020,Du2020,Anand2021a,Leyton-Ortega2021,Demers2018,Hu2019,Zeng2019,Zoufal2019,verdon2019quantum,Huang2020,Situ2020,Coyle2021,Liu2021,rudolph2020generation}.

\subsection{Born Machines as Implicit Generative Models} 
\label{ssec:born_machines}

By directly exploiting Born's probabilistic interpretation of quantum wave functions~\cite{born1926quantenmechanik}, it is possible to model the probability distribution of classical data using a pure quantum state. Such models are referred to as Born machines~\cite{cheng2018information}. We are particularly interested in Born machines for which the quantum state is obtained via a parameterised quantum circuit (as opposed to, say, a continuous time Hamiltonian evolution). These are known as quantum circuit Born machines (QCBMs)~\cite{Liu2018,Benedetti2019a}.

The use of QCBMs as generative models is in large part motivated by their expressiveness. Indeed, it is now well established that Born machines have greater expressive power than classical models, including neural networks~\cite{Du2020} and partially matrix product states~\cite{Glasser2019} (see also~\cite{Coyle2020}). This means, in particular, that QCBMs can efficiently represent certain distributions which are classically intractable to simulate (e.g.,~\cite{Bremmer2016,Boixo2018,Bouland2019}). These include those recently used in a demonstration of quantum supremacy~\cite{Arute2019}.

Let us consider a binary vector $\boldsymbol{x}\in\{0,1\}^n$, with $n$ the number of qubits. A QCBM takes a product state $|0\rangle^{\otimes n}$ as input and evolves it into a normalised output state $|\Psi(\boldsymbol{\theta})\rangle$ via a parameterised quantum circuit $U(\boldsymbol{\theta})$. One can generate $n$-bit strings according to
\begin{equation}
    \boldsymbol{x}\sim q_{\boldsymbol{\theta}}(\boldsymbol{x})= |\langle \boldsymbol{x}|\Psi(\boldsymbol{\theta})\rangle|^2 ,
\label{born}
\end{equation}
where $|\boldsymbol{x}\rangle$ are computational basis states; sampling from this distribution then consists of a simple measurement. Since we only have access to $\boldsymbol{x}\sim q_{\boldsymbol{\theta}}(\boldsymbol{x})$ and not the probabilities, $q_{\boldsymbol{\theta}}(\boldsymbol{x})$ themselves, the Born machine can be regarded as an implicit generative model. We~consider parameterised quantum circuits $U(\boldsymbol{\theta})$ of the form
\begin{equation}
    U(\boldsymbol{\theta}) = \prod_{i=1}^D  W_iU_i(\theta_i) ,
\label{pqc}
\end{equation}
where $\{W_i\}_{i=1}^D$ is a set of fixed unitaries, $\{U_i(\theta_i)\}_{i=1}^D$ is a set of parameterised unitaries, and $D$ is the depth of circuit. We also assume that $\smash{U_i(\theta_i) = e^{-i\theta_i V_i}}$ are rotations through angles $\theta_i$, generated by Hermitian operators $V_i$ with eigenvalues $\pm1$. In this case, one can compute partial derivatives of $q_{\boldsymbol{\theta}}(\boldsymbol{x})$ using the parameter-shift rule~\cite{Mitarai2018}, which reads
\begin{equation}
    \partial_{\theta_i} q_{\boldsymbol{\theta}}(\boldsymbol{x}) =  q_{\boldsymbol{\theta}_{i}^{+}}(\boldsymbol{x}) - q_{\boldsymbol{\theta}_{i}^{-}}(\boldsymbol{x}) ,
\label{param_shift}
\end{equation}
where $\boldsymbol{\theta}_{{i}^{\pm}} = \boldsymbol{\theta}\pm \frac{\pi}{4} \boldsymbol{e}_i$, with $\boldsymbol{e}_i$ a unit vector in the $i^{\text{th}}$ direction.
More generally, this formula allows one to express the 
first-order partial derivative of an expectation of a function $h$ as
\begin{equation}
    \partial_{\theta_i} \mathbb{E}_{\boldsymbol{x}\sim q_{\boldsymbol{\theta}}(\boldsymbol{x})} [h(\boldsymbol{x})] = \mathbb{E}_{\boldsymbol{x}\sim q_{\boldsymbol{\theta}_{i}^{+}}(\boldsymbol{x})} [h(\boldsymbol{x})] - \mathbb{E}_{\boldsymbol{x}\sim q_{\boldsymbol{\theta}_{i}^{-}}(\boldsymbol{x})} [h(\boldsymbol{x})] .
\end{equation}

The major challenge in using any implicit generative model is designing a suitable objective function. As noted before, one cannot compute $q_{\boldsymbol{\theta}}(\boldsymbol{x})$ directly, and thus valid objectives can only involve statistical quantities (e.g., expectations) which can be efficiently computed using samples. 
For generative models based on QCBMs, various objectives have been proposed, including moment-matching, maximum mean discrepancy, Stein and Sinkhorn divergences, and adversarial objectives based on the Kullback-Leibler divergence. In this paper, we propose a more general class of objective functions -- $f$-divergences -- for training QCBMs.

\subsection{Adversarial Generative Modelling with \texorpdfstring{$f$}{f}-Divergences}
\label{ssec:f_divergences}

Let $f:(0,\infty)\rightarrow\mathbb{R}$ be a convex function with $f(1)=0$ and strict convexity at 1. Suppose that $p(\boldsymbol{x})=0$ whenever $q_{\boldsymbol{\theta}}(\boldsymbol{x})=0$. The $f$-divergence, or Csisz\'{a}r divergence~\mbox{\cite{Csiszar1967information, Ali1966general}},  between $q_{\boldsymbol{\theta}}$ and $p$ is defined as
\begin{equation}
    D_f ( p \| q_{\boldsymbol{\theta}} ) = \mathbb{E}_{\boldsymbol{x}\sim q_{\boldsymbol{\theta}}(\boldsymbol{x})} \left[ f\left( \frac{p(\boldsymbol{x})}{q_{\boldsymbol{\theta}}(\boldsymbol{x})} \right) \right] .
\label{eq:f_div}
\end{equation}
Suppose instead that $q_{\boldsymbol{\theta}}(\boldsymbol{x})=0$ whenever $p(\boldsymbol{x})=0$. Then the $f$-divergence can be written as 
\begin{equation}
    D_f ( p \| q_{\boldsymbol{\theta}} ) = \mathbb{E}_{\boldsymbol{x}\sim p(\boldsymbol{x})} \left[ f^{*}\left( \frac{q_{\boldsymbol{\theta}}(\boldsymbol{x})}{p(\boldsymbol{x})} \right) \right] .
\label{eq:fdiv_conj}
\end{equation}
where the conjugate function is defined as $f^{*}(r)=r f(1/r)$ (not to be confused with the Fenchel conjugate). In what follows, we will generally prefer this formulation, as it leads to simpler expressions. 

The function $f$ is called the \emph{generator} of the divergence. For different choices of $f$, one obtains well-known divergences such as TV, KL, and $\chi^2$. In this paper, we investigate the effect of this choice on the training of a QCBM. To ensure a fair comparison, we assume that the generators are standardised and normalised such that $f'(1)=0$ and $f''(1)=1$~\cite{amari2009alpha}.
This ensures that $D_f(p \| q_{\boldsymbol{\theta}}) \geq 0$ with equality if and only if $p \equiv q_{\boldsymbol{\theta}}$, even if $p$ and $q_{\boldsymbol{\theta}}$ are unnormalised. Note that one can normalise and standardise 

We minimise the $f$-divergence using gradient-based methods. We thus require the derivative of $D_f$ with respect to $\boldsymbol{\theta}_i$. Using the chain and the parameter-shift rules, it is straightforward to compute
\begin{eqnarray}
    \partial_{\boldsymbol{\theta}_i} D_f(p \| q_{\boldsymbol{\theta}} )
    &=& \sum_{\boldsymbol{x}} p(\boldsymbol{x}) \partial_{\theta_i} f^{*}\left( \frac{q_{\boldsymbol{\theta}} (\boldsymbol{x})}{p (\boldsymbol{x})} \right) \\
    &=& \sum_{\boldsymbol{x}} p(\boldsymbol{x}) f^{*\prime} \left( \frac{q_{\boldsymbol{\theta}} (\boldsymbol{x})}{p (\boldsymbol{x})} \right) \frac{1}{p(\boldsymbol{x})}  \partial_{\theta_i} q_{\boldsymbol{\theta}}(\boldsymbol{x}) \\
    &=& \sum_{\boldsymbol{x}} f^{*\prime} \left( \frac{q_{\boldsymbol{\theta}} (\boldsymbol{x})}{p (\boldsymbol{x})} \right) \left( q_{\boldsymbol{\theta}_i^+}(\boldsymbol{x}) -  q_{\boldsymbol{\theta}_i^-}(\boldsymbol{x}) \right)  \\
    &=& \mathbb{E}_{\boldsymbol{x} \sim q_{\boldsymbol{\theta}_i^+}(\boldsymbol{x})} \left[ f^{*\prime} \left( \frac{q_{\boldsymbol{\theta}} (\boldsymbol{x})}{p (\boldsymbol{x})} \right) \right] - \mathbb{E}_{\boldsymbol{x} \sim q_{\boldsymbol{\theta}_i^-}(\boldsymbol{x})} \left[ f^{*\prime} \left( \frac{q_{\boldsymbol{\theta}} (\boldsymbol{x})}{p (\boldsymbol{x})} \right) \right] . 
\label{eq:fdiv_gradient}
\end{eqnarray}

\begin{figure}[t]
\centering
    \begin{subfigure}[b]{0.34\linewidth}
        \centering
        \includegraphics[width=\textwidth]{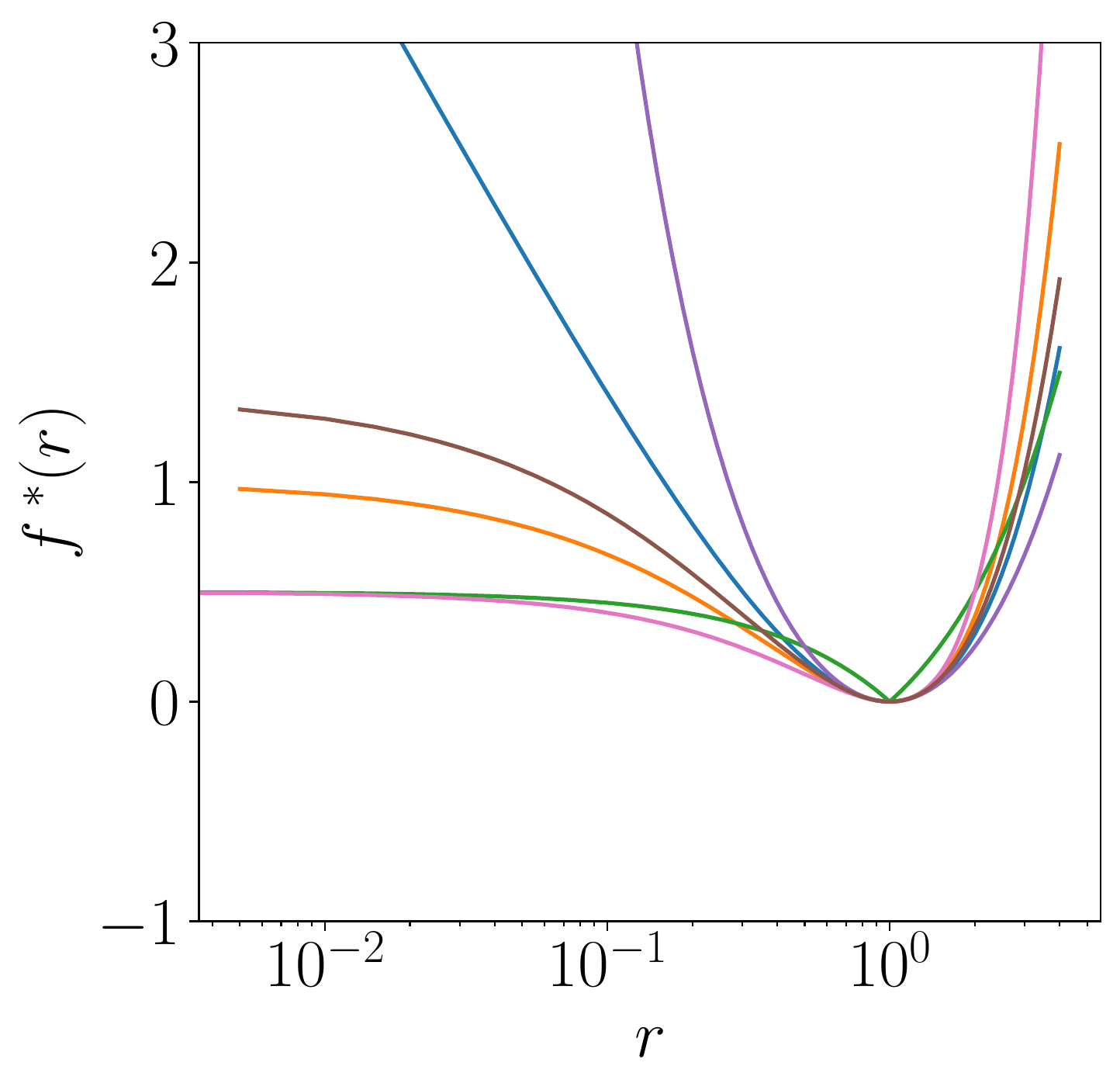}
    \end{subfigure}
    \hspace{8mm}
    \begin{subfigure}[b]{0.34\linewidth}
        \centering
        \includegraphics[width=\textwidth]{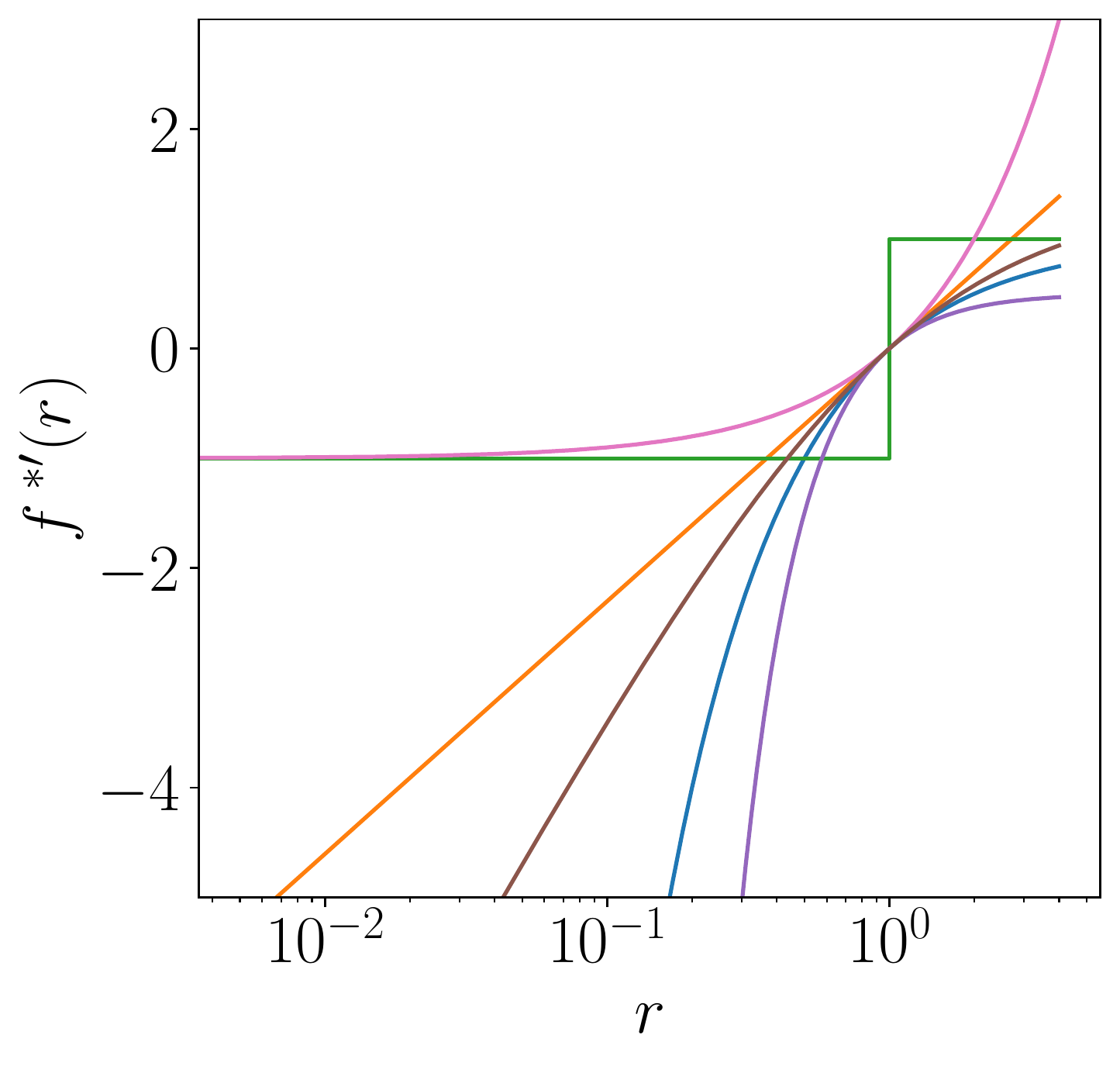}
    \end{subfigure}
    \hspace{8mm}
    \begin{subfigure}[b]{0.14\linewidth}
        \centering
        \raisebox{.5\height}{
            \includegraphics[width=\textwidth]{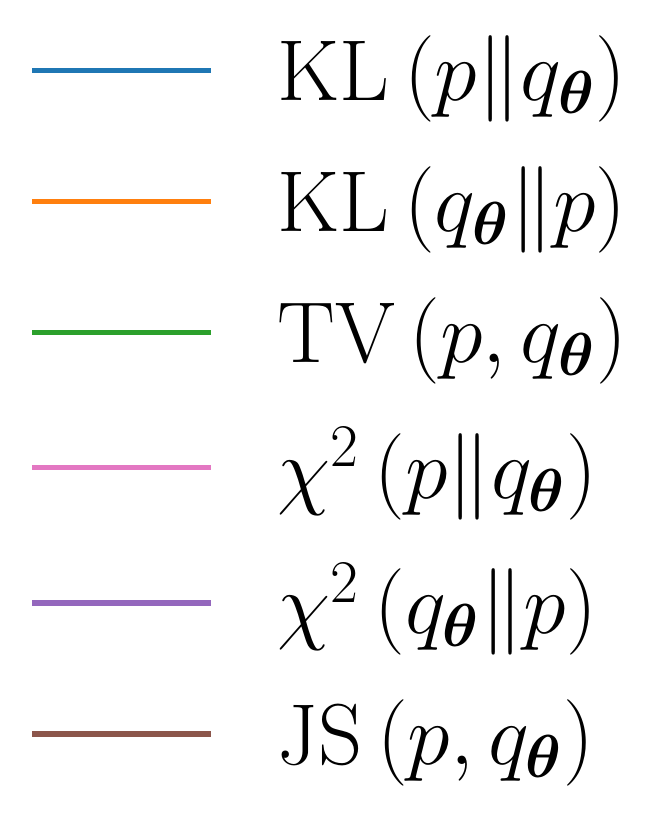}
        }
    \end{subfigure}
    \caption{Conjugate $f^{*}$ (left panel) and derivative $f^{*\prime}$ (right panel) of the generator $f$ for several $f$-divergences. All generators have been standardised with $f^{\prime}(1)=0$ and normalised with $f^{\prime\prime}(1)=1$, except for TV.}
    \label{fig:generators}
\end{figure}

We summarise some well-known $f$-divergences, the conjugates of their generators, and their parameter-shift rules, in Tables \ref{t:f_divs} and \ref{t:f_divs_sym}. We also plot some of the conjugate generators in Figure \ref{fig:generators}.

Returning to Equation~\eqref{eq:fdiv_gradient}, it is clear that the problem of computing the gradient reduces to that of estimating the probability ratio $r(\boldsymbol{x}) = \frac{q_{\boldsymbol{\theta}}(\boldsymbol{x})}{p(\boldsymbol{x})}$. We choose to define $r(\boldsymbol{x})$ in this way since it is more natural when one is interested with writing the $f$-divergence in terms of $f^*$, as we do here. Note that in some literature the ratio is defined in the reverse manner by switching the probabilities. We can estimate the probability ratio from the output of a binary classifier~\cite{Sugiyama2012}. Suppose we assign samples $\boldsymbol{x}$$\sim$$q_{\boldsymbol{\theta}}(\boldsymbol{x})$ to one class, and samples $\boldsymbol{x}$$\sim$$p(\boldsymbol{x})$ to another class. Suppose, in addition, that one has access to an exact binary classifier $d_{*}(\boldsymbol{x})$, which outputs the probability that the sample $\boldsymbol{x}$ originated from $q_{\boldsymbol{\theta}}(\boldsymbol{x})$. Then,  assuming uniform prior probabilities for the two classes, it is straightforward to show via Bayes' theorem that (see Section \ref{ssec:born_machines} in~\cite{Mohamed2017})
\begin{align}
    r(\boldsymbol{x}) =\frac{d_{*}(\boldsymbol{x})}{1-d_{*}(\boldsymbol{x})} .
\label{eq:est_ratio}
\end{align}

In practice, we do not have access to the exact classifier $d_{*}(\boldsymbol{x})$. However, under the assumption that we can efficiently sample from both distributions, we can train a classifier $d_{\boldsymbol{\phi}}(\boldsymbol{x})$, parameterised by $\boldsymbol{\phi}$, to distinguish between the two distributions. One can use any proper scoring rule to train the classifier~\cite{Mohamed2017}. A typical choice is the negative cross entropy, given by
\begin{equation}
    \mathcal{L}(\boldsymbol{\phi} ; \boldsymbol{\theta}) = -\mathbb{E}_{\boldsymbol{x} \sim q_{\boldsymbol{\theta}}(\boldsymbol{x})} \left[ \log{d_{\boldsymbol{\phi}}(\boldsymbol{x})} \right] - \mathbb{E}_{\boldsymbol{x} \sim p(\boldsymbol{x})} \left[ \log(1 - d_{\boldsymbol{\phi}}(\boldsymbol{x})) \right] .
\label{eq:negative_cross_entropy}
\end{equation}
The classifier seeks to minimise this objective, which corresponds to low classification errors. We emphasise that, in this objective, $\boldsymbol{\theta}$ is fixed at the current QCBM parameters. The resulting classifier approximates the probability ratio for the current QCBM as
\begin{equation}
    r(\boldsymbol{x}) \approx \frac{d_{\boldsymbol{\phi}}(\boldsymbol{x})}{1- d_{\boldsymbol{\phi}}(\boldsymbol{x})} .
\label{eq:ratio_approx}
\end{equation}
This can be plugged into Equation~\eqref{eq:fdiv_gradient} to approximate the gradient. With this in mind, we define the cost function for the QCBM as
\begin{equation}
    \mathcal{J}(\boldsymbol{\theta}; \boldsymbol{\phi}) = \mathbb{E}_{\boldsymbol{x}\sim q_{\boldsymbol{\theta}}(\boldsymbol{x})} \left[f^{*\prime}\left( \frac{d_{\boldsymbol{\phi}}(\boldsymbol{x})}{1- d_{\boldsymbol{\phi}}(\boldsymbol{x})} \right) \right] ,
\label{eq:f_div_approx}
\end{equation}
where now the parameters of the classifier $\boldsymbol{\phi}$ are fixed and the argument of the expectation value is independent of $\boldsymbol{\theta}$. The adversarial generative modelling can be regarded as the following optimisation problem 
\begin{eqnarray}
    \boldsymbol{\theta}^{*}&=&\argmin_{\boldsymbol{\theta}}\mathcal{J}(\boldsymbol{\theta} ; \boldsymbol{\phi}) , \\
    \boldsymbol{\phi}^{*}&=&\argmin_{\boldsymbol{\phi}}\mathcal{L}(\boldsymbol{\phi}; \boldsymbol{\theta}) ,
\end{eqnarray}
where the required expectation values are estimated from samples.
In principle, the classifier can be trained to optimality in order to provide the best possible ratio for the generative model. Alternatively, the two objective functions can be optimised in tandem, using alternating gradient descent steps or a two-timescale gradient descent scheme~\cite{Heusel2017}.

\begin{table}[t]
\small
\renewcommand{\arraystretch}{2.5}
\scalebox{1}{
\begin{tabular}{c|c|c|c}
\textbf{\emph{f}-Divergence} & \textbf{Definition} & \boldmath{$f^{*}$} & \textbf{Parameter-Shift}\\
\hline
total variation & 
$\text{TV}(p, q_{\boldsymbol{\theta}}) = \tfrac{1}{2}\sum |p(\boldsymbol{x}) - q_{\boldsymbol{\theta}}(\boldsymbol{x})|$ &
$\tfrac{1}{2}|r - 1|$ & 
$\tfrac{1}{2} \mathbb{E}_{q_{\boldsymbol{\theta}^{+}}}\left[\text{sgn}(r(\boldsymbol{x})-1)\right] - \tfrac{1}{2} \mathbb{E}_{q_{\boldsymbol{\theta}^{-}}}\left[\text{sgn}(r(\boldsymbol{x})-1)\right]$ \\
squared Hellinger & 
$\text{H}^2(p, q_{\boldsymbol{\theta}}) = \sum {( \sqrt{p(\boldsymbol{x})} - \sqrt{q_{\boldsymbol{\theta}}(\boldsymbol{x})})}^2$ & 
$2(\sqrt{r}-1)^2$ &
$-2\mathbb{E}_{q_{\boldsymbol{\theta}^{+}}}\left[\tfrac{1}{\sqrt{r(\boldsymbol{x})}}\right] + 2\mathbb{E}_{q_{\boldsymbol{\theta}^{-}}}\left[\tfrac{1}{\sqrt{r(\boldsymbol{x})}}\right]$ \\
\makecell{Kullback-Leibler\\(type I, forward)}  & 
$\text{KL}(p \| q_{\boldsymbol{\theta}}) = \mathbb{E}_p \left[ \log{\frac{p(\boldsymbol{x})}{q_{\boldsymbol{\theta}}(\boldsymbol{x})}} \right]$ & 
$-\log{r} + r - 1$ &
$-\mathbb{E}_{q_{\boldsymbol{\theta}^{+}}}\left[\tfrac{1}{r(\boldsymbol{x})}\right] + \mathbb{E}_{q_{\boldsymbol{\theta}^{-}}}\left[\tfrac{1}{r(\boldsymbol{x})}\right]$ \\
\makecell{Kullback-Leibler\\(type I, reverse)} & 
$\text{KL}(q_{\boldsymbol{\theta}} \|p) =  \mathbb{E}_{q_{\boldsymbol{\theta}}} \left[ \log{\frac{q_{\boldsymbol{\theta}}(\boldsymbol{x})}{p(\boldsymbol{x})}} \right]$ & 
$r \log{r} - r + 1$ & 
$\mathbb{E}_{q_{\boldsymbol{\theta}^{+}}}\left[\log r(\boldsymbol{x})\right] - \mathbb{E}_{q_{\boldsymbol{\theta}^{-}}}\left[\log r(\boldsymbol{x})\right]$ \\
\makecell{Kullback-Leibler\\(type II, forward)} & 
$\text{KL}(p \| \frac{p+q_{\boldsymbol{\theta}}}{2} ) = \mathbb{E}_p \left[ \log{\frac{2p(\boldsymbol{x})}{p(\boldsymbol{x})+q_{\boldsymbol{\theta}}(\boldsymbol{x})}} \right]$ & 
$4 \log{\frac{2}{r+1}} + 2(r-1)$ & 
$-4\mathbb{E}_{q_{\boldsymbol{\theta}^{+}}}\left[\tfrac{1}{r(\boldsymbol{x})+1}\right] + 4\mathbb{E}_{q_{\boldsymbol{\theta}^{-}}}\left[\tfrac{1}{r(\boldsymbol{x})+1}\right]$ \\
\makecell{Kullback-Leibler\\(type II, reverse)} & 
$\text{KL}(q_{\boldsymbol{\theta}}\| \frac{p+q_{\boldsymbol{\theta}}}{2} ) = \mathbb{E}_{q_{\boldsymbol{\theta}}} \left[ \log{\frac{2q_{\boldsymbol{\theta}}(\boldsymbol{x})}{p(\boldsymbol{x})+q_{\boldsymbol{\theta}}(\boldsymbol{x})}} \right]$ & 
$4r \log{\frac{2r}{r+1}} + 2(1-r)$ & 
\makecell{ $4\mathbb{E}_{q_{\boldsymbol{\theta}^{+}}}\left[\log\frac{r(\boldsymbol{x})}{r(\boldsymbol{x})+1} + \frac{1}{r(\boldsymbol{x})+1}\right] $ \\ \qquad\qquad $- 4\mathbb{E}_{q_{\boldsymbol{\theta}^{-}}}\left[\log\frac{r(\boldsymbol{x})}{r(\boldsymbol{x})+1} + \frac{1}{r(\boldsymbol{x})+1}\right]$} \\
\makecell{Pearson\\(forward)} & 
$\chi^2(p\|q_{\boldsymbol{\theta}}) = \sum \frac{{(p(\boldsymbol{x}) - q_{\boldsymbol{\theta}}(\boldsymbol{x}))}^2}{p(\boldsymbol{x})}$ & 
$\frac{(r-1)^2}{2}$ & 
$\mathbb{E}_{q_{\boldsymbol{\theta}^{+}}}\left[r(\boldsymbol{x})\right] - \mathbb{E}_{q_{\boldsymbol{\theta}^{-}}}\left[r(\boldsymbol{x})\right]$ \\
\makecell{Pearson\\(reverse)} & 
$\chi^2(q_{\boldsymbol{\theta}} \| p) = \sum \frac{{(p(\boldsymbol{x}) - q_{\boldsymbol{\theta}}(\boldsymbol{x}))}^2}{q_{\boldsymbol{\theta}}(\boldsymbol{x})}$ & $\frac{(r-1)^2}{2r}$ & 
$-\tfrac{1}{2}\mathbb{E}_{q_{\boldsymbol{\theta}^{+}}}\left[\tfrac{1}{r(\boldsymbol{x})^2}\right] + \tfrac{1}{2}\mathbb{E}_{q_{\boldsymbol{\theta}^{-}}}\left[\tfrac{1}{r(\boldsymbol{x})^2}\right]$ \\
\end{tabular}
}
\caption{A summary of well-known $f$-divergences, including the definition, the conjugate of the generator $f^{*}$, and the corresponding parameter-shift rule in terms of the ratio $r(\boldsymbol{x}) = \frac{q_{\boldsymbol{\theta}}(\boldsymbol{x})}{p(\boldsymbol{x})}$.
The $\|$ symbol indicates that the divergence is asymmetric, while a comma indicates that it is symmetric. Interestingly, one can construct symmetric $f$-divergences for every asymmetric one (see Table \ref{t:f_divs_sym}).}
\label{t:f_divs}
\end{table}

\begin{table}[t]
\small
\renewcommand{\arraystretch}{2.5}
\scalebox{1}{
\begin{tabular}{c|c|c} 
\textbf{\emph{f}-Divergence} & \textbf{Definition} & \textbf{Parameter-Shift}\\ 
\hline
\makecell{symmetric Kullback-Leibler\\(type I, Jeffrey)} & $\text{J}(p,q_{\boldsymbol{\theta}}) = \text{KL}(p\|q_{\boldsymbol{\theta}}) + \text{KL}(q_{\boldsymbol{\theta}}\|p)$ & $\tfrac{1}{2}\mathbb{E}_{q_{\boldsymbol{\theta}^{+}}}\left[\log{r(\boldsymbol{x})} - \tfrac{1}{r(\boldsymbol{x})}\right] - \tfrac{1}{2}\mathbb{E}_{q_{\boldsymbol{\theta}^{-}}}\left[\log{r(\boldsymbol{x})} - \tfrac{1}{r(\boldsymbol{x})}\right]$ \\
\makecell{symmetric Kullback-Leibler\\(type II, Jensen-Shannon)} & $\text{JS}(p,q_{\boldsymbol{\theta}}) = \text{KL}(p\| \frac{p+q_{\boldsymbol{\theta}}}{2} ) + \text{KL}(q_{\boldsymbol{\theta}}\| \frac{p+q_{\boldsymbol{\theta}}}{2} )$ & $2\mathbb{E}_{q_{\boldsymbol{\theta}^{+}}}\left[\log{\frac{r(\boldsymbol{x})}{1+r(\boldsymbol{x})}}\right] - 2\mathbb{E}_{q_{\boldsymbol{\theta}^{-}}}\left[\log{\frac{r(\boldsymbol{x})}{1+r(\boldsymbol{x})}}\right]$ \\
\makecell{symmetric Pearson} & $\bar{\chi}^2(p, q_{\boldsymbol{\theta}}) = \chi^2(p\|q_{\boldsymbol{\theta}}) + \chi^2(q_{\boldsymbol{\theta}} \| p)$ & $\tfrac{1}{4}\mathbb{E}_{q_{\boldsymbol{\theta}^{+}}}\left[2r(\boldsymbol{x}) - \frac{1}{r(\boldsymbol{x})^2}\right] - \tfrac{1}{4}\mathbb{E}_{q_{\boldsymbol{\theta}^{-}}}\left[2r(\boldsymbol{x}) - \frac{1}{r(\boldsymbol{x})^2}\right]$ \\
\end{tabular}
}
\caption{A summary of the symmetric $f$-divergences corresponding to some well-known asymmetric $f$-divergences, including the definition, and the parameter-shift rule.}
\label{t:f_divs_sym}
\end{table}

\section{Training Heuristics} 
\label{sec:training_strategies}

\subsection{Switching \texorpdfstring{$f$}{f}-Divergences} 
\label{ssec:switching_f_divergences}

In this Section, we describe a heuristic for dynamically switching between $f$-divergences throughout the training process of our generative model (specifically the QCBM).

To motivate this heuristic, we examine how $D_f(p \| q_{\boldsymbol{\theta}})$ varies with respect to values of $
\smash{r(\boldsymbol{x})=\frac{q_{\boldsymbol{\theta}}(\boldsymbol{x})}{p(\boldsymbol{x})}}$. We begin by noting that all $f$-divergences which can be standardised agree on the divergence between nearby distributions~\cite{csiszar2004information}, but can otherwise exhibit very different behaviours. In particular, we focus on their initial rates of convergence. 

One may rationalise the different rates of convergence for each divergence at the beginning of training by considering the following argument~\cite{Goodfellow2014, Mohamed2017, Uehara2016generative}.
Consider $n$ qubits, such that there are $2^n$ different values of $r(\boldsymbol{x})$. For a successful training, all these values need to converge towards $1$ (which implies our goal that $q_{\boldsymbol{\theta}} \equiv p$). Now suppose we were to estimate the divergence in Equation~\eqref{eq:fdiv_conj} using a set of samples from the target distribution $\smash{\boldsymbol{x}_1,\dots,\boldsymbol{x}_m\stackrel{\text{i.i.d.}}{\sim} p(\boldsymbol{x})}$. At the beginning of training, $q_{\boldsymbol{\theta}}$ is initialised at random and is therefore expected to be far from the target. This means that $q_{\boldsymbol{\theta}} (\boldsymbol{x}_i) \ll p(\boldsymbol{x}_i)$ for most of the samples. In other words, at the beginning of training most of the samples yield probability ratios $r(\boldsymbol{x}_i) \ll 1$. 

It is evident from the left panel of Figure~\ref{fig:generators} that some divergences, including TV, vary slowly in the region where $r \ll 1$, and are therefore more liable to saturation in the initial stages of training. Other divergences, such as forward KL and reverse $\chi^2$, generate strong gradients in this region.  In the limiting case where $p$ and $q_{\boldsymbol{\theta}}$ have disjoint supports, TV and JS saturate, whereas forward KL diverges~\cite{Arjovsky2017wasserstein}. This problem is well known within the context of training generative adversarial networks; since an idealised formulation optimises JS, several alternative cost functions have been proposed to mitigate its slow initial convergence~\cite{Goodfellow2014, Nowozin2016, Uehara2016generative, Arjovsky2017wasserstein}. 

Though we can only apply this logic to the particular regime where $p$ and $q_{\boldsymbol{\theta}}$ are far apart, it is also evident from Figure~\ref{fig:generators} that the $f$-divergences exhibit a wide diversity of behaviours throughout most of training. We propose to exploit this with the following heuristic. At every optimisation step, we choose an $f$-divergence for each direction in parameter space that generates the highest gradient in said direction. This requires no additional quantum circuit evaluations since we only need to evaluate Equation~\eqref{eq:fdiv_gradient} for the different generators. Concretely, the heuristic can be written as follows. For each step, to update parameter $\boldsymbol{\theta}_i$, we choose the $f$-divergence labelled $j$, $D_{f_j}$, which obeys
\begin{equation} 
\label{eqn:f_switch_criterion}
    \lvert \partial_{\boldsymbol{\theta}_i} D_{f_j} \rvert > \lvert \partial_{\boldsymbol{\theta}_i} D_{f_k} \rvert \qquad \forall k \in \mathcal{F} .
\end{equation}

For simplicity, in this paper, we restrict the set $\mathcal{F}$ to only contain those $f$-divergences illustrated in Figure~\ref{fig:generators}.
We call this heuristic \emph{$f$-switch}.

\subsection{Local Cost Functions} 
\label{ssec:local_cost_funcs}

In this Section, we outline an alternative heuristic for training the QCBM, based on introducing locality into the cost function. Let us briefly provide some motivation for this approach. One of the most fundamental challenges associated with hybrid quantum-classical algorithms is the \emph{barren plateau} phenomenon, whereby the gradient of the cost function vanishes exponentially in the number of qubits~\cite{McClean2018,Grant2019,Arrasmith2020,Marrero2020,Patti2020,Uvarov2021,Arrasmith2021,Cerezo2021,Holmes2021,Larocca2021,Wang2021}. This phenomenon can arise due to deep unstructured ans\"atze~\cite{McClean2018}, large entanglement~\cite{Marrero2020,Patti2020}, high levels of noise~\cite{Wang2021}, and global cost functions~\cite{Cerezo2021,Uvarov2021}. As such, it is a rather general phenomenon in many quantum machine learning applications, including generative models. In the presence of barren plateaus, exponential precision (i.e., an exponential number of samples) is required in order to resolve against finite sampling noise and determine a minimising direction in the cost function landscape. Since the standard objective of quantum algorithms is to achieve a polynomial scaling in the system size (as opposed to the exponential scaling of classical algorithms), barren plateaus can destroy any hope of a variational quantum algorithm achieving quantum advantage.
 
Although, in this paper, we do not directly analyse the emergence of barren plateaus in the QCBM, we are nonetheless motivated by existing results on barren plateaus. We focus, in particular, on the connection between barren plateaus and global cost functions (i.e., cost functions defined in terms of global observables), given that such cost functions naturally arise in hybrid quantum-classical generative models. The connection between trainability and locality was first established by Cerezo et al.~\cite{Cerezo2021}, who proved that cost functions defined in terms of global observables exhibit barren plateaus for \emph{all} circuit depths in circuits composed of random two-qubit gates which act on alternating pairs of qubits (i.e., blocks forming local 2-designs). Meanwhile, local cost functions do not exhibit barren plateaus for shallow circuits; in this case, cost function gradients vanish at worst polynomially in the number of qubits.  

On the basis of this result, there is clear motivation to seek a local cost function (i.e., a cost function defined in terms of local observables) for the hybrid quantum-classical generative model introduced in Section \ref{ssec:f_divergences}. We now attempt to make some progress towards this goal. 

We write $q^{i}_{\boldsymbol{\theta}}(\boldsymbol{x}_i)$ to denote the marginal distribution of the $i^{\text{th}}$ element of the bit-string $\boldsymbol{x} = (\boldsymbol{x}_1,\dots,\boldsymbol{x}_n)$. Using Jensen's inequality on Equation~\eqref{eq:fdiv_conj}, it can be shown that the $f$-divergence between joint distributions is larger than the $f$-divergence between marginal distributions. Thus, we have
\begin{equation}
    D_f( p(\boldsymbol{x}) \| q_{\boldsymbol{\theta}}(\boldsymbol{x}) )  \geq \frac{1}{n} \sum_{i=1}^n D_f( p^i(\boldsymbol{x}_i) \| q^i_{\boldsymbol{\theta}}(\boldsymbol{x}_i) ) .
\label{eq:fdiv_ineq}
\end{equation}

Our heuristic consists of minimising the right-hand side of this inequality. Even though this is a lower-bound to the original cost, it is a fully local cost function. Later, we show how to generalise this approach allowing for a trade off between trainability and accuracy. We call this heuristic \emph{f-local}.

Let us show the difference between the global cost function (left-hand side of the inequality) and the local cost function (right-hand side) by means of an example. For ease of exposition, we assume in this discussion that the $f$-divergence of interest is the reverse KL with generator $f^*(r) = r \log r - r + 1$. We emphasise, however, that the methodology is generic to any $f$-divergence.
We begin by rewriting the expression in Equation~\eqref{born} as
\begin{equation}
    q_{\boldsymbol{\theta}}(\boldsymbol{x})= \langle \boldsymbol{0}| U^{\dagger}(\boldsymbol{\theta}) H_{\boldsymbol{x}} U(\boldsymbol{\theta})|\boldsymbol{0}\rangle ,
\end{equation}
where we have defined $H_{\boldsymbol{x}} := |\boldsymbol{x}\rangle \langle \boldsymbol{x}|$. We can thus write the reverse KL in the form of a generic cost function (see, e.g.,~\cite{Cerezo2020a}) as
\begin{align}
    \text{KL}(q_{\boldsymbol{\theta}}\|p) = \sum_{\boldsymbol{x}} q_{\boldsymbol{\theta}}(\boldsymbol{x})\log \frac{q_{\boldsymbol{\theta}}(\boldsymbol{x})}{p(\boldsymbol{x})}=\sum_{\boldsymbol{x}} g_{\boldsymbol{x}}\bigg(\langle\boldsymbol{0}| U^{\dagger}(\boldsymbol{\theta}) H_{\boldsymbol{x}}U(\boldsymbol{\theta}) |\boldsymbol{0}\rangle\bigg) ,
\label{eq:reverseKL}
\end{align}
where we define $\smash{g_{\boldsymbol{x}}(q_{\boldsymbol{\theta}}) := q_{\boldsymbol{\theta}}\log \frac{{q}_{\boldsymbol{\theta}}}{p(\boldsymbol{x})}}$. This cost function is clearly global, since the observables, $H_{\boldsymbol{x}}$, act on all qubits.

Now, rewriting Equation~\eqref{eq:reverseKL} in terms of the adversarial approximation in Equation~\eqref{eq:f_div_approx}, we have
\begin{align}
    \mathcal{J}(\boldsymbol{\theta}; \boldsymbol{\phi}) =\sum_{\boldsymbol{x}} q_{\boldsymbol{\theta}}(\boldsymbol{x})\logit \left(d_{\boldsymbol{\phi}}(\boldsymbol{x})\right)=\sum_{\boldsymbol{x}}  h_{\boldsymbol{x}}\bigg(\langle\boldsymbol{0}| U^{\dagger}(\boldsymbol{\theta}) H_{\boldsymbol{x}}U(\boldsymbol{\theta}) |\boldsymbol{0}\rangle\bigg) ,
\label{global}
\end{align}
where $\smash{h_{\boldsymbol{x}}(q_{\boldsymbol{\theta}}) := q_{\boldsymbol{\theta}}\logit(d_{\boldsymbol{\phi}}(\boldsymbol{x}))}$, and $\logit(y) := \log \frac{y}{1-y}$.
It is interesting to note that the global observable $H_{\boldsymbol{x}}$ only enters into $h_{\boldsymbol{x}}(q_{\boldsymbol{\theta}})$ via the first term, namely $q_{\boldsymbol{\theta}}(\boldsymbol{x})$. It is arguable, however, that the second term in  $h_{\boldsymbol{x}}(q_{\boldsymbol{\theta}})$, namely $\logit( d_{\boldsymbol{\phi}}(\boldsymbol{x}) )$ should also be regarded as a global quantity. 

We now consider the fully local cost function in the right-hand side of Equation~\eqref{eq:fdiv_ineq}. Applying the adversarial approximation to each of the $n$ probability ratios, the QCBM objective is
\begin{align}
    \mathcal{J}^{L}(\boldsymbol{\theta} ; \boldsymbol{\phi}) 
    =\frac{1}{n} \sum_{\vphantom{\boldsymbol{x}_i\in\{0,1\}}i=1}^n \sum_{\boldsymbol{x}_i\in\{0,1\}}^{\vphantom{n}}  q^{i}_{\boldsymbol{\theta}}(\boldsymbol{x}_i)\logit \left( d^{i}_{\boldsymbol{\phi}}(\boldsymbol{x}_i) \right) = \frac{1}{n}\sum_{\vphantom{\boldsymbol{x}_i\in\{0,1\}}i=1}^n\sum_{\boldsymbol{x}_i\in\{0,1\}} h_{\boldsymbol{x}_i}^{L}\bigg(\langle\boldsymbol{0}| U^{\dagger}(\boldsymbol{\theta}) H_{\boldsymbol{x}_i}^{L}U(\boldsymbol{\theta}) |\boldsymbol{0}\rangle\bigg) ,
\end{align}
where we have replaced the global observable $H_{\boldsymbol{x}}$ in Equation~\eqref{global} by the set of local observables 
\begin{equation}
    H_{\boldsymbol{x}_i}^{L} = |\boldsymbol{x}\rangle \langle \boldsymbol{x}|_i\otimes \mathds{1}_{\tilde{i}} .
\end{equation}
Here, $|\boldsymbol{x}\rangle\langle \boldsymbol{x}|_{i}$ is a projector on the computational basis for qubit $i$, and $\mathds{1}_{\tilde{i}}$ denotes the identity on all qubits except qubit $i$. We have also replaced the `global' function $\smash{h_{\boldsymbol{x}}(q_{\boldsymbol{\theta}})}$ in Equation~\eqref{global} by the set of local functions 
\begin{equation}
    h_{\boldsymbol{x}_i}^{L}(p^{i}_{\boldsymbol{\theta}}) = q^{i}_{\boldsymbol{\theta}} \logit \left(d_{\boldsymbol{\phi}}^{i}(\boldsymbol{x}_i)\right) .
\end{equation}
Here, $\smash{\{d_{\boldsymbol{\phi}}^{i}(\boldsymbol{x}_i)\}_{i=1}^n}$ is a set of $n$ `local' classifiers, which act only on the marginal distribution corresponding to the $i^{\text{th}}$ qubit. That is to say, $\smash{d_{\boldsymbol{\phi}}^{i}}$ are trained to distinguish between samples $\smash{\boldsymbol{x}_i\sim q^{i}_{\boldsymbol{\theta}}(\boldsymbol{x}_i)}$ and samples $\smash{\boldsymbol{x}_i\sim p^{i}(\boldsymbol{x}_i)}$. One may ask why it is not sufficient to simply make only the observable, $H_{\boldsymbol{x}}$, local as is done in other literature addressing the barren plateau problem~\cite{Cerezo2021}. In our case, it turns out that if one does not \emph{also} make the functions $h_{\boldsymbol{x}}$ local, in other words by keeping the classifier `global', the cost function becomes intractable to compute due to a need to explicitly compute joint probabilities from the circuit, $q_{\boldsymbol{\theta}}$. This hints at the subtlety that appears when attempting to address barren plateaus in generative modelling, that does not necessarily exist in other variational algorithms.

We are, of course, interested in whether the local cost function is faithful to the original cost function. Recall that we are minimising the lower bound in Equation~\eqref{eq:fdiv_ineq}. It is clear that, if the local cost function is minimised, so that $D_f(q^{i}_{\boldsymbol{\theta}}\|p^{i})=0$ for all $i\in\{1,\dots,n\}$, and all of the marginals coincide, there is still no guarantee that the joint distributions will be identical.
This observation suggests that, while this cost function may be more trainable than the original cost function on account of its locality, it may also be significantly less accurate. 
In an attempt to remedy this, we can instead consider a more general $k$-local cost function which acts on subsets of $k$ qubits. 
In particular, by defining $\boldsymbol{x}_{i:j} := (\boldsymbol{x}_i,\dots,\boldsymbol{x}_j)$, we can introduce
\begin{align}
    \mathcal{J}^{L(k)}(\boldsymbol{\theta}; \boldsymbol{\phi}) &= \frac{1}{n-k+1} \sum_{\vphantom{\boldsymbol{x}_{i:i+k-1}\in\{0,1\}^k}i=1}^{n-k+1} \sum_{\boldsymbol{x}_{i:i+k-1}\in\{0,1\}^k}^{\vphantom{n}} q_{\boldsymbol{\theta}}^{i:i+k-1}(\boldsymbol{x}_{i:i+k-1})\logit\left(d^{i:i+k-1}_{\boldsymbol{\phi}}(\boldsymbol{x}_{i:i+k-1})\right) 
\label{LK2}  \\
    &= \frac{1}{n-k+1} \sum_{\vphantom{\boldsymbol{x}_{i:i+k-1}\in\{0,1\}^k}i=1}^{n-k+1} \sum_{\boldsymbol{x}_{i:i+k-1}\in\{0,1\}^k}^{\vphantom{n}} h_{\boldsymbol{x}_{i:i+k-1}}^{L(k)}\bigg(\langle\boldsymbol{0}| U^{\dagger}(\boldsymbol{\theta}) H_{\boldsymbol{x}_{i:i+k-1}}^{L(k)}U(\boldsymbol{\theta}) |\boldsymbol{0}\rangle\bigg) ,
\label{LK3} 
\end{align}
where 
\begin{align}
    H^{L(k)}_{\boldsymbol{x}_{i:i+k-1}} &= |\boldsymbol{x}\rangle\langle \boldsymbol{x}|_{i:i+k-1}\otimes \mathds{1}_{\widetilde{i:i+k-1}} ,
\label{eq:obs_def} \\
    h^{L(k)}_{\boldsymbol{x}_{i:i+k-1}}(q^{i:i+k-1}_{\boldsymbol{\theta}}) &= q^{i:i+k-1}_{\boldsymbol{\theta}}\logit \left(d_{\boldsymbol{\phi}}^{i:i+k-1}(\boldsymbol{x}_{i:i+k-1})\right) ,
\label{eq:func_def}
\end{align}
and where $\{d_{\boldsymbol{\phi}}^{i:i+k-1}(\boldsymbol{x}_{i:i+k-1})\}_{i=1}^{n-k+1}$ is a set of $n-k+1$ `$k$-local' classifiers, defined in an obvious fashion. This $k$-local cost function now approximates the sum of the reverse KL between the $k$-marginals (of neighbouring qubits) of the target distribution $p(\boldsymbol{x})$, and the variational distribution $q_{\boldsymbol{\theta}}(\boldsymbol{x})$. 

Arguing as before, it is clear that the $k$-local cost function will admit additional global minima in comparison to the global cost function for any $1\leq k<n$. In particular, when the $k$-local cost function is minimised, the $k$-nearest neighbour marginals of $p(\boldsymbol{x})$ and $q_{\boldsymbol{\theta}}(\boldsymbol{x})$ coincide. One can expect, however, that as the value of $k$ is increased, not only will the number of additional minima decrease, but the disparity between the joint distributions of the target and the model at these global minima will decrease. 
This suggests that in order to achieve a `sweet spot' between trainability and accuracy, a reasonably approach is to start by optimising the $k$-local cost function with a small value of $k$ (promoting trainability), before iteratively increasing the value of $k$ (promoting accuracy) until $k=n$, thus recovering the global cost function. 

We should remark that, while for ease of notation we have defined the $k$-local cost function in terms of marginals with respect to neighbouring qubits $(\boldsymbol{x}_{i},\dots,\boldsymbol{x}_{i+k-1})$, one can in theory choose any sets of qubits of size at most $k$ (e.g., nearest neighbours, all possible combinations, and randomly sampled). In general, for a fixed value of $k$, this choice will influence the accuracy of the objective function, as well as its computational cost, and should be made on a case-by-case basis on the basis of the available computational resources.

\section{Numerical Results} 
\label{sec:results}

\begin{figure}[t]
\centering
    \includegraphics[width=.8\linewidth]{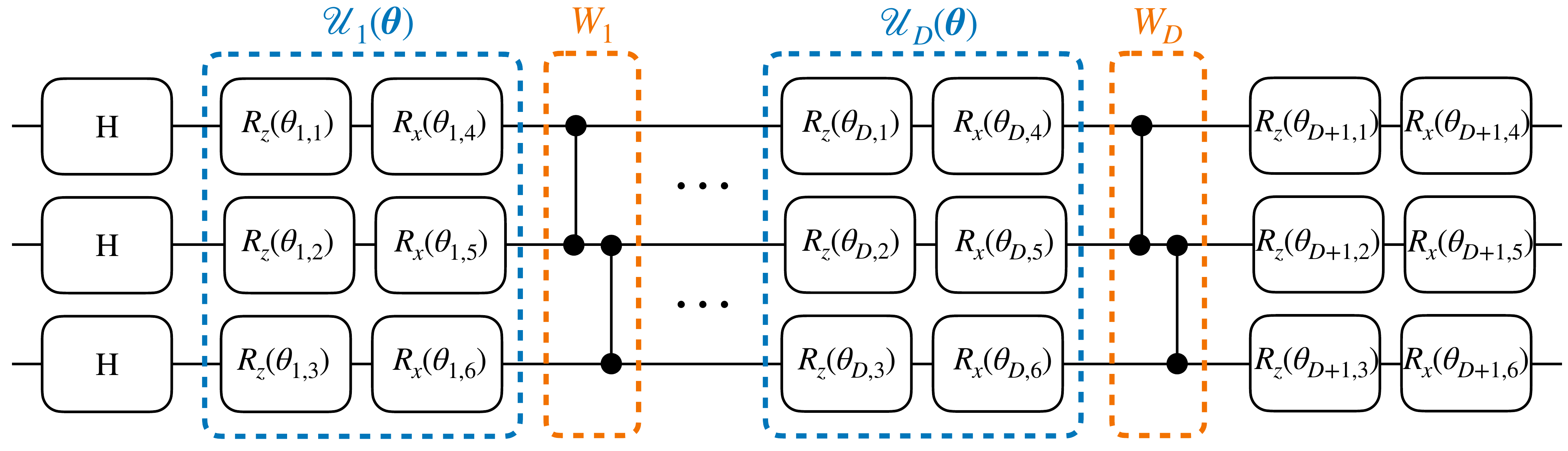}
    \caption{The ansatz employed in numerical simulations (shown for three qubits). The ansatz consists of $D$ alternating layers of single qubit gates and entangling gates. The single qubit layers consists of two single qubit rotations, one around the $z$ axis and one around the $x$ axis. The entangling layer is composed of a ladder of CZ gates. There is an additional layer of Hadamard gates prior to the first layer, and an additional layer of single qubit rotations after the final layer. The total number of parameters in a circuit of depth $D$ is given by $n_p = n(2D + 2)$, where $n$ is the number of qubits.}
   \label{fig:ansatz}
\end{figure}

In this Section, we present numerical results to illustrate the performance of the training heuristics proposed in Section \ref{sec:training_strategies}. Throughout this Section, we utilise a QCBM composed of alternating layers of single qubit gates and entangling gates (see Figure \ref{fig:ansatz}). We implement the quantum circuit using \texttt{pytket}~\cite{Sivarajah2020} and execute the simulations with Qiskit~\cite{Qiskit}. The parameters of the QCBM are updated using stochastic gradient descent with a constant learning rate, which is tuned to each of the simulations.

\begin{figure}[t]
\centering
\captionsetup[subfigure]{justification=centering,skip=0pt}
    \subfloat[3 qubits]{\includegraphics[width=.42\linewidth]{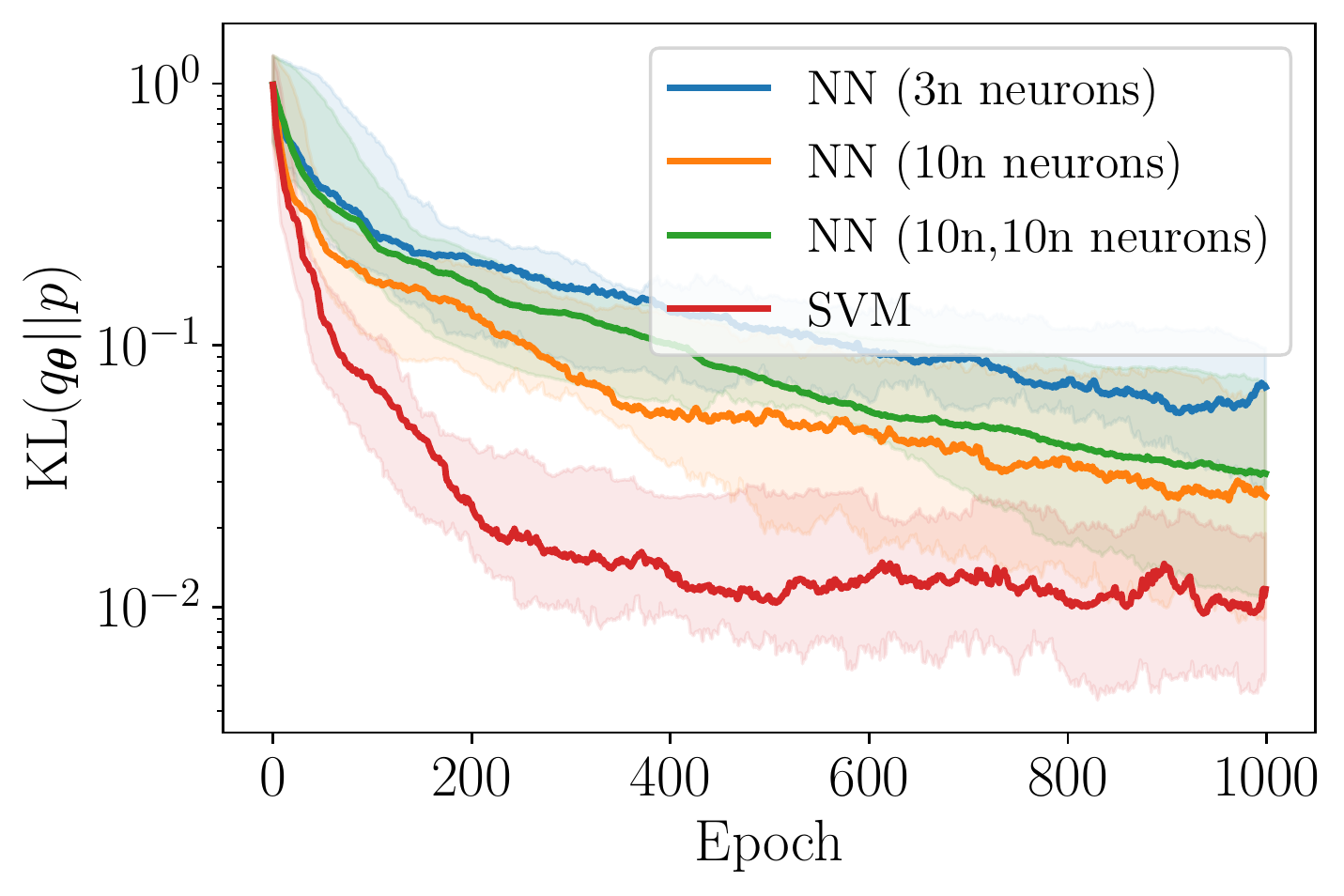}\label{fig:classifier_1a}}
    \hspace{8mm}
    \subfloat[4 qubits]{\includegraphics[width=.42\linewidth]{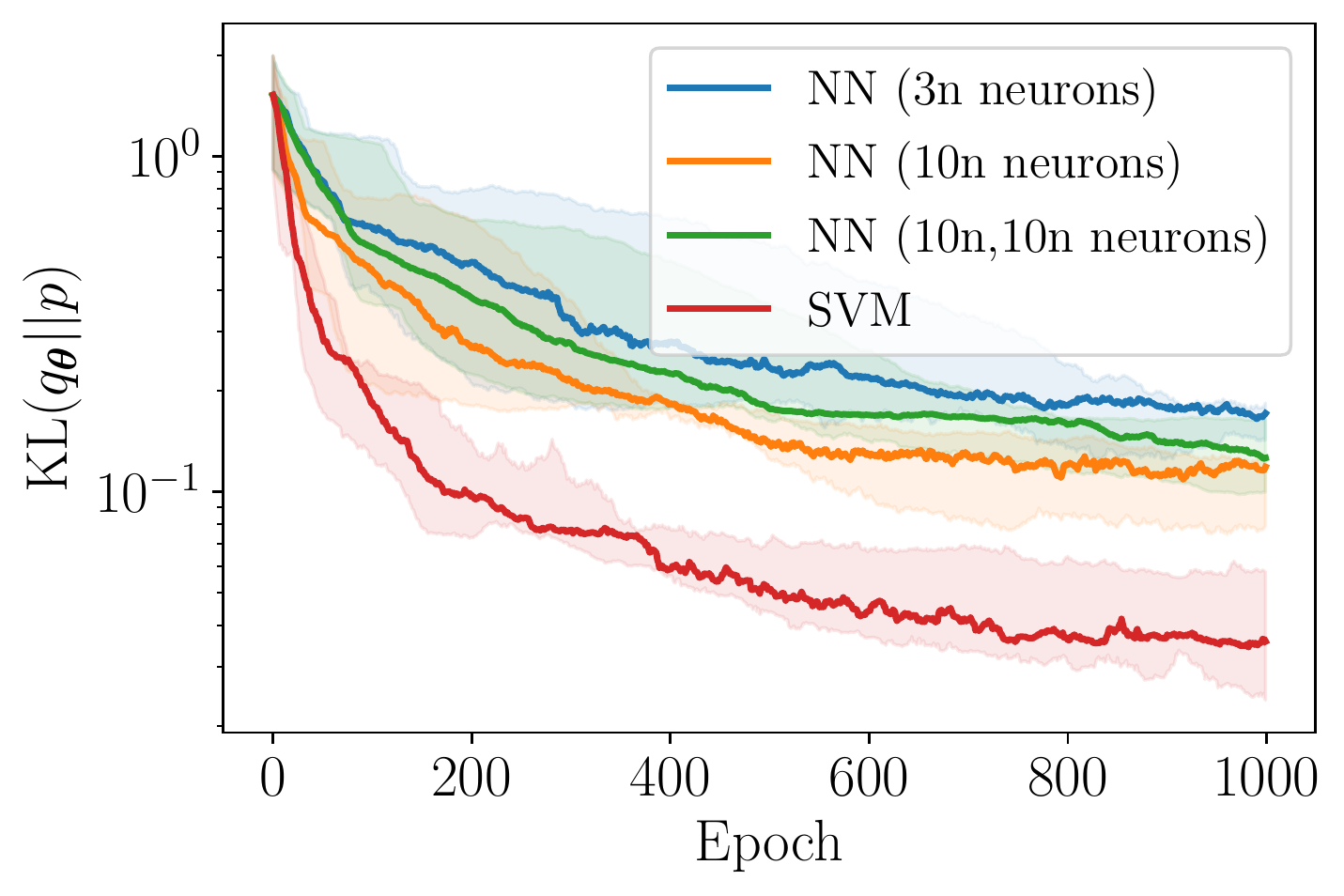}\label{fig:classifier_1b}}
    \caption{Training performance of the QCBM in illustrative 3 qubit and 4 qubit experiments using 4 different classifiers. The classifiers are trained using 500 samples. We plot the bootstrapped median (solid line), as well as 90\% confidence intervals (shaded).}
\label{fig:classifier_comparison}
\end{figure}

Regarding the classical component of the adversarial generative model (i.e., the binary classifier), we use either a fully connected feed-forward neural network with ReLU neurons (NN), or a support vector machine with RBF kernel (SVM). Indeed, one rather surprising byproduct of our numerical investigation is that the training performance of the adversarial generative model could be improved, at times significantly, by using a SVM in place of a NN for this component (see Figure \ref{fig:classifier_comparison}). This, in itself, should be of some interest to practitioners. Not only can SVMs be faster to train, but they depend on significantly fewer hyper-parameters than NNs, whose performance is often highly dependent on careful tuning of the number of hidden layers, the number of neurons in each hidden layer, the learning rate, the batch size, etc. While we do not suggest that SVMs will always outperform NNs in this setting, this does indicate that SVMs may represent a viable alternative. We implement the NNs using PyTorch~\cite{PyTorch}, while the SVMs are implemented with \texttt{scikit-learn}~\cite{scikit-learn}. The particular hyper-parameters used in each simulation are specified below.

In the majority of our numerical simulations, we consider a QCBM with $3$ qubits. This corresponds to a discrete target  distribution $p$ which takes $2^3$ values. We generally also assume that the target distribution corresponds to a particular instantiation of the QCBM, for a fixed number of layers, $D_p$. By varying the number of layers, $D_{q_{\boldsymbol{\theta}}}$ used to train the generative model, we can then investigate different parameterisation regimes of interest. In the case that the number of layers used to generate the target is greater than the number of layers used in the model ($D_p>D_{q_{\boldsymbol{\theta}}}$), the model is under-parameterised (or severely under-parameterised). Meanwhile, when the number of layers used to generate the target and the number of layers used in the model are equal ($D_p = D_{q_{\boldsymbol{\theta}}}$), the model is said to be exactly parameterised. In these cases, a solution to the learning problem is guaranteed to exist: there exists $\boldsymbol{\theta}=\boldsymbol{\theta}_0$ such that $p\equiv q_{\boldsymbol{\theta_0}}$. Finally, when the number of layers used to generate the target is less than the number used in the model ($D_p<D_{q_{\boldsymbol{\theta}}}$), the model is over-parameterised (or severely over-parameterised). We provide a more precise definition of these different cases, as applied to our numerics, in Table \ref{tab:parameterisations}.

\begin{table}[t]
\small
\renewcommand{\arraystretch}{2.}
\scalebox{1}{
\begin{tabular}{p{.20\linewidth} |p{.14\linewidth}p{.14\linewidth}p{.14\linewidth}p{.14\linewidth}p{.14\linewidth}}
& Severely Over \newline Parameterised \newline (OO) &  Over \newline Parameterised \newline (O) & Exactly \newline Parameterised \newline (E) & Under \newline Parameterised \newline (U)  & Severely Under \newline Parameterised \newline (UU)  \\ 
\hline
Number of parameters\newline (layers) used to \newline generate the target $p$  & 12 parameters \newline (1 layer)  & 12 parameters \newline (1 layer) & 12 parameters \newline (1 layer)   & 30 parameters \newline (4 layers)      & 30 parameters \newline (4 layers) \\
Number of parameters\newline (layers) used for the \newline model $q_{\boldsymbol{\theta}}$  & 30 parameters \newline (4 layers) & 24 parameters \newline (3 layers) & 12 parameters \newline (1 layer) & 18 parameters \newline (2 layers) & 12 parameters \newline (1 layer) \\
\end{tabular}
}
\caption{The different parameterisation regimes used in the 3 qubit numerical simulations.}
\label{tab:parameterisations}
\end{table}

For each of the settings (i.e., choice of circuit depth for the target and model, choice of heuristic, number of qubits) explored, we train the generative model using nine independent parameter initialisations. We then use a bootstrapping procedure to provide a more robust estimate of the median cost at each training epoch. We first take samples of size nine from the outcome of the nine independent experiments, 10,000 times with replacement. We then compute the median cost across each set of samples to obtain a distribution of 10,000 medians. Using this distribution, we compute the median and obtain error bars from the $5^\text{th}$ and $95^\text{th}$ percentiles, corresponding to a $90\%$ confidence interval.

\subsection{Switching \texorpdfstring{$f$}{f}-Divergences}

We begin by considering the performance of the heuristic introduced in Section \ref{ssec:switching_f_divergences}. The $f$-divergences that can be standardised locally behave as KL to second order~\cite{csiszar2004information}. Notably, TV cannot be standardised; indeed, it is straightforward to show that TV provides an upper bound for all other $f$-divergences with $f^{\prime \prime}(1)=1$ in this regime.
For this reason, we evaluate both the exact TV and the exact KL to measure performance.

We begin by reporting the results obtained using an exact classifier, for each of the parameterisation regimes given in Table \ref{tab:parameterisations}. The generator is trained using $1000$ samples per iteration. The results are given in Table \ref{tab:f_switch_results}.

\begin{figure}[t]
    \begin{subfigure}[b]{0.34\linewidth}
        \includegraphics[width=\textwidth]{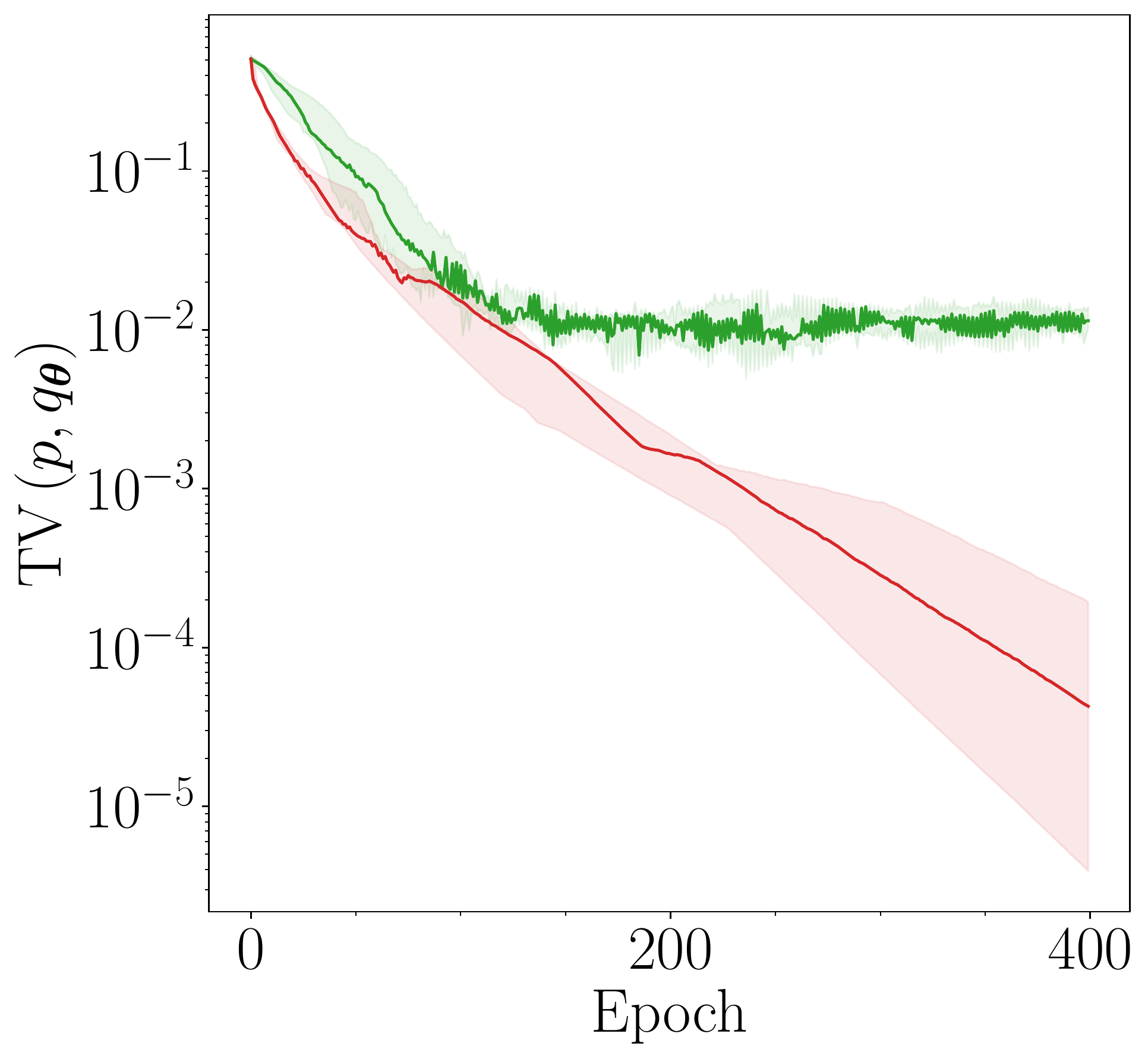}
    \end{subfigure}
    \hspace{8mm}
    \begin{subfigure}[b]{0.34\linewidth}
        \includegraphics[width=\textwidth]{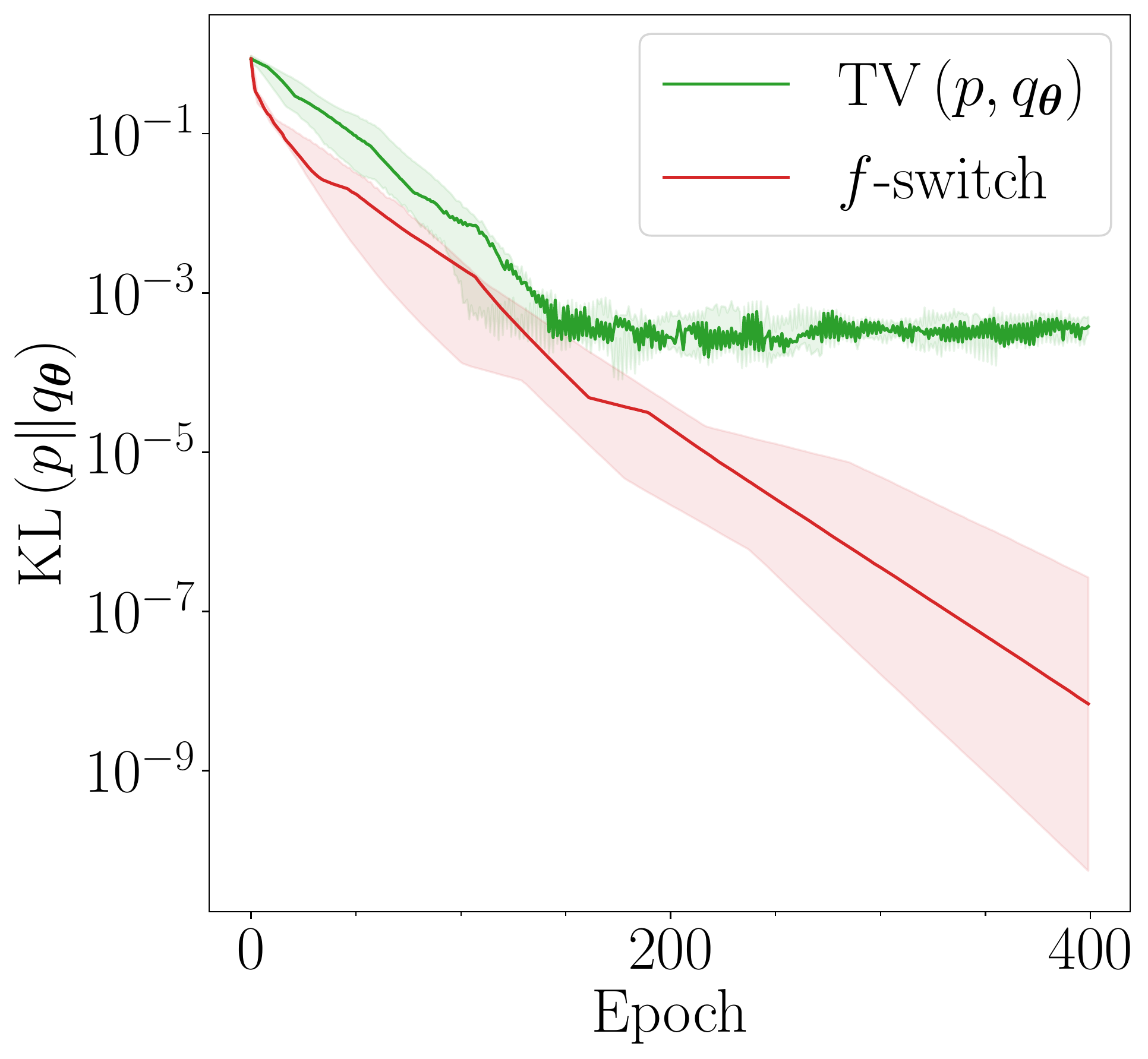}
    \end{subfigure}
\caption{Performance of the QCBM trained using the TV (green) and the $f$-divergence heuristic (red) for 3 qubits in the severely over-parameterised case OO(12,30), using an exact classifier. We show the bootstrapped median (solid line) and 90\% confidence intervals (shaded) of the TV (left) and the KL (right).}
\label{fig:OO_median_exact}
\end{figure}

\begin{figure}[t]
    \begin{subfigure}[b]{0.34\linewidth}
        \includegraphics[width=\textwidth]{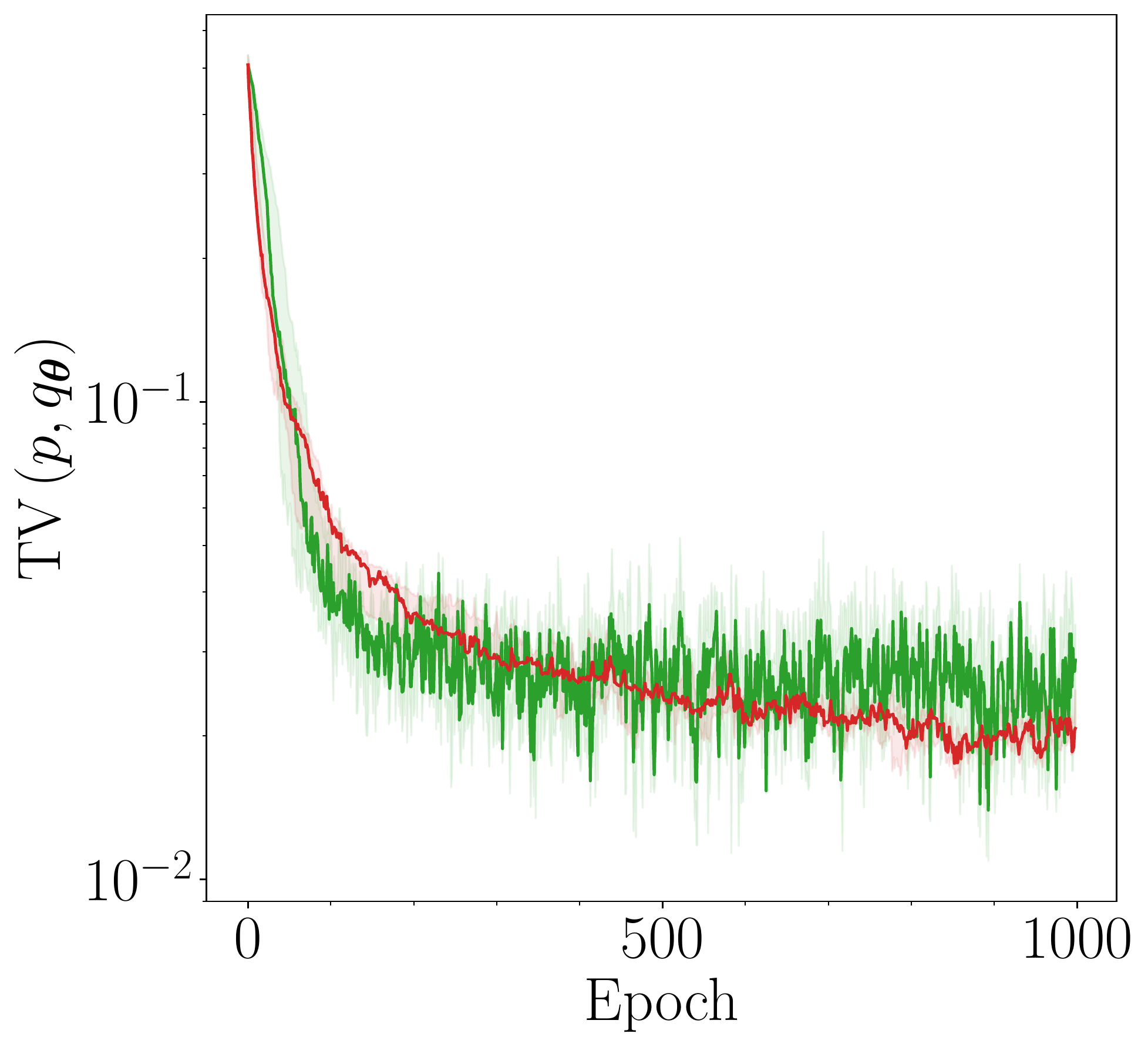}
    \end{subfigure}
    \hspace{8mm}
    \begin{subfigure}[b]{0.34\linewidth}
        \includegraphics[width=\textwidth]{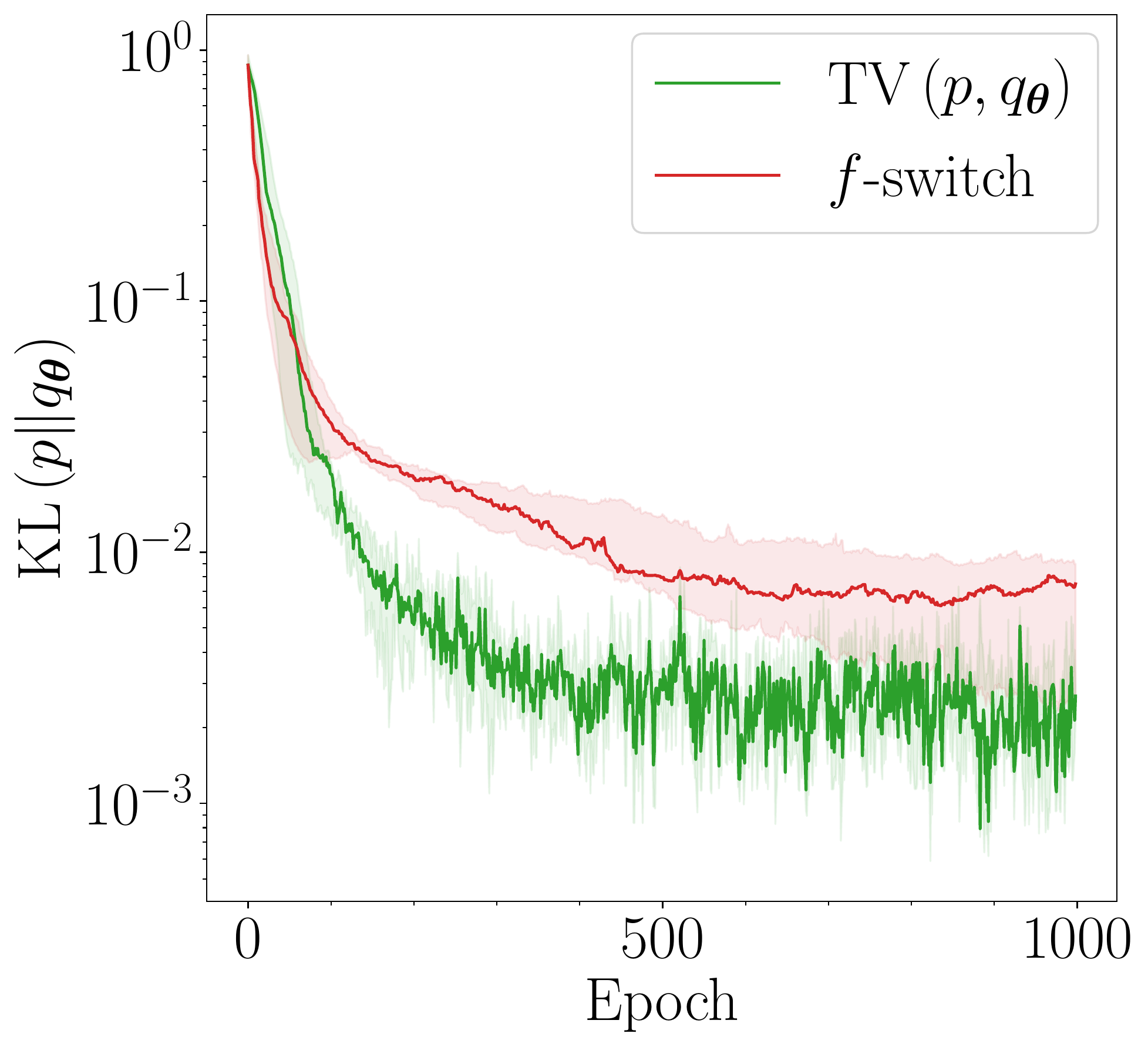}
    \end{subfigure}
\caption{Performance of the QCBM training using the TV (green) and the $f$-divergence heuristic (red) for 3 qubits in the severely over-parameterised case OO(12,30), using a trained SVM classifier. We show the bootstrapped median (solid line) and 90\% confidence intervals (shaded) of both the TV (left) and the KL (right).}
\label{fig:OO_median_trained}
\end{figure}

Our results indicate that the heuristic is able to outperform TV when the QCBM is (severely) over-parameterised. This may be due to the extra degrees of freedom in the model. These allow for more discrepancies between the loss landscapes of the $f$-divergences, which the heuristic is able to exploit. In Figures \ref{fig:OO_median_exact} and \ref{fig:OO_median_trained}, we provide a more detailed illustration of the training performance of the $f$-switch heuristic in this regime. Figure \ref{fig:OO_median_exact} corresponds to an exact classifier: in this case, use of the heuristic significantly improves the convergence of the QCBM. Figure \ref{fig:OO_median_trained} corresponds to a trained classifier, trained on $1000$ samples per iteration: in this case, use of the heuristic can lead to marginal performance improvements with respect to TV (left-hand figure).
The remaining results in this Section are all reported for an exact classifier.

\begin{table}[t]
\footnotesize
\renewcommand{\arraystretch}{2.}
\scalebox{1}{
\begin{tabular}{p{.08\linewidth}|p{.09\linewidth}|p{.185\linewidth}|p{.155\linewidth}|p{.135\linewidth}|p{.135\linewidth}|p{.135\linewidth}} 
    $\boldsymbol{D_f}$ \newline \textbf{evaluated} & $\boldsymbol{D_f}$ \textbf{used \newline in training} & \textbf{OO \newline (12, 30)} & \textbf{O \newline(12, 24)} & \textbf{E \newline(12, 12)} & \textbf{U \newline(30, 18)} & \textbf{UU \newline(30, 12)} \\
    \hline
    TV & TV & $\left(1.12 \substack{+0.45 \\ -0.28}\right)\times 10^{-2}$ & $\left(8.4 \substack{+1.2 \\ -1.0}\right)\times 10^{-3}$ & $\left(1.0\substack{+1.51 \\ -0.12}\right)\times 10^{-2}$ & $\left(1.06 \substack{+0.26 \\ -0.23}\right)\times 10^{-2}$ & $\left(1.4\substack{+2.4 \\ -0.7}\right)\times 10^{-2}$ \\
    TV & $f$-switch & $\boldmath{\left(0.6\substack{+3.8 \\ -0.5}\right)\times 10^{-5}} *$ & $\boldmath{\left(2.5\substack{+2.5 \\ -2.1}\right)\times 10^{-3}} *$ & $\left(3.1 \substack{+1.8 \\ -1.9}\right)\times 10^{-2}$ & $\left(0.65 \substack{+0.27 \\ -0.51}\right)\times 10^{-2}$ & $\left(1.8 \substack{+2.9 \\ -0.9}\right)\times 10^{-2}$ \\
    KL & TV & $\left(3.5 \substack{+2.1 \\ -1.3}\right)\times 10^{-4}$ & $\left(2.0 \substack{+0.6 \\ -0.4}\right)\times 10^{-4}$ & $\left(2.6 \substack{+14.8 \\ -2.3}\right)\times 10^{-3}$ & $\left(3.7 \substack{+1.7 \\ -92.6}\right)\times 10^{-4}$ & $\left(0.6\substack{+24.3 \\ -0.4}\right)\times 10^{-3}$ \\
    KL & $f$-switch & $\boldmath{\left(0.0182\substack{+1.383 \\ -0.012}\right)\times 10^{-8}}$ & $\left(1.8\substack{+20.9 \\ -1.7}\right)\times 10^{-5} *$ & $\left(3.5 \substack{+9.1 \\ -2.0}\right)\times 10^{-3}$ & $\left(2.4 \substack{+1.6 \\ -2.4}\right)\times 10^{-4}$ & $\left(1.8 \substack{+4.3 \\ -1.5}\right)\times 10^{-3}$ \\
\end{tabular}
}
\caption{Performance of the QCBM trained using the TV and the $f$-divergence heuristic for $3$ qubits in over-, under-, and exactly parameterised regimes. We show the bootstrapped median of the TV (top two rows) and the KL (bottom two rows) after 500 epochs. The asterisk (*) on some of the experiments indicates that the cost is still converging. The bold indicates the regimes where $f$-switch significantly outperforms the other methods.}
\label{tab:f_switch_results}
\end{table}

\begin{figure}[t]
    \begin{subfigure}[b]{0.34\linewidth}
        \includegraphics[width=\textwidth]{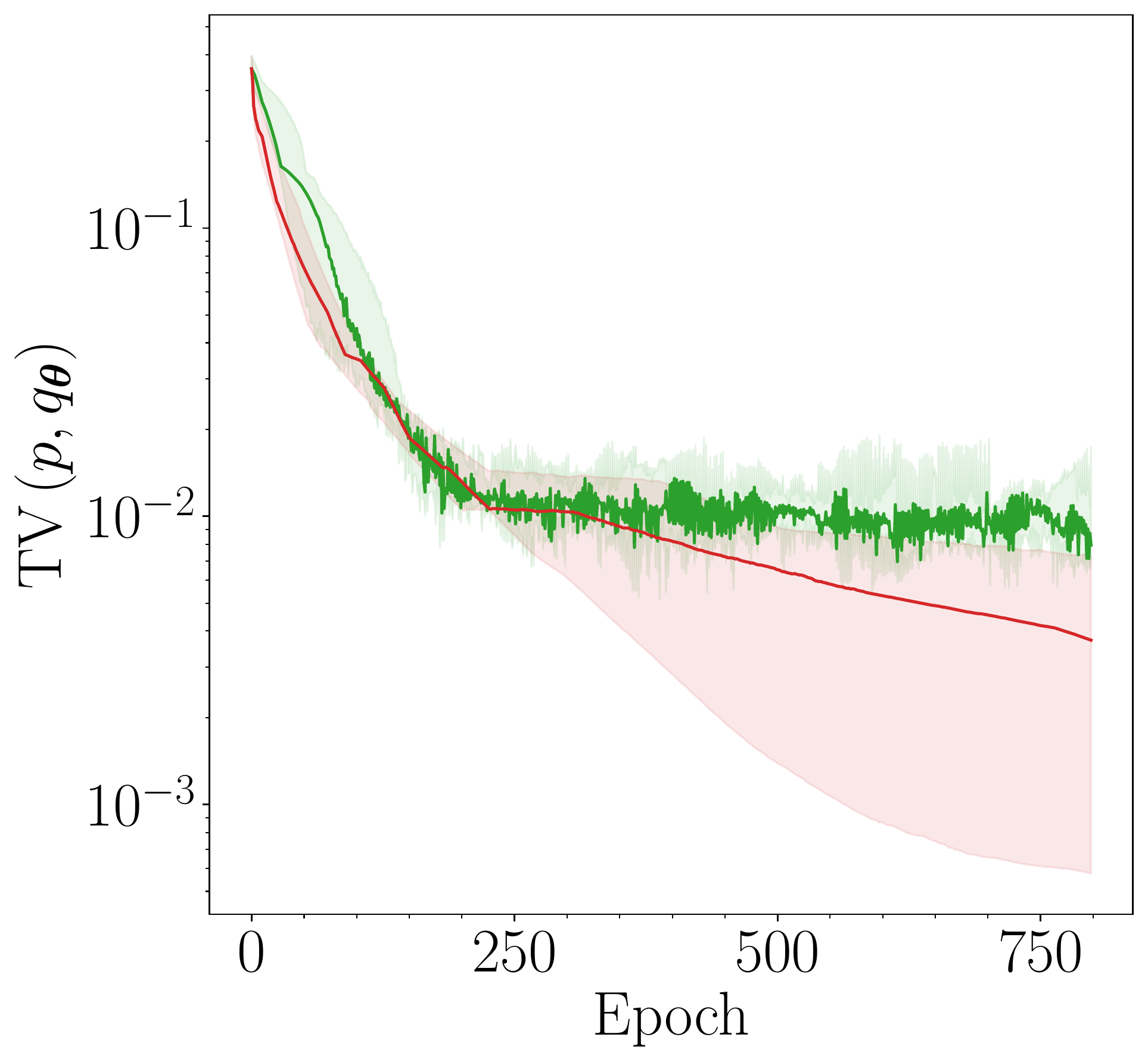}
    \end{subfigure}
    \hspace{8mm}
    \begin{subfigure}[b]{0.34\linewidth}
        \includegraphics[width=\textwidth]{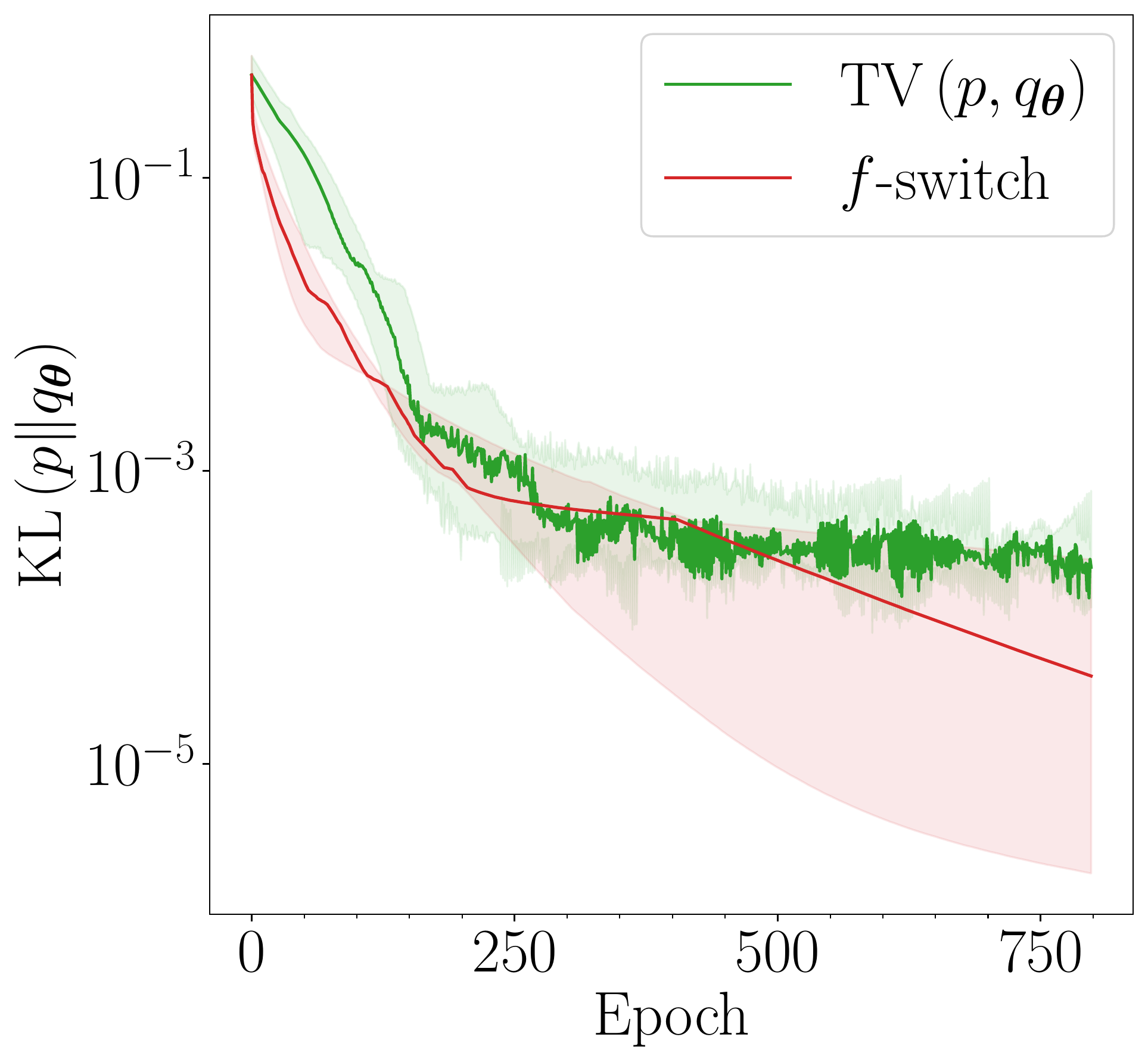}
    \end{subfigure}
    \caption{Performance of the QCBM training using the TV (green) and the $f$-divergence heuristic (red) for 3 qubits in the under-parameterised case U(30,18). We show the bootstrapped median (solid line) and 90\% confidence intervals (shaded) of both the TV (left) and the KL (right).}
    \label{fig:U_all_initial_configs}
\end{figure}

\begin{figure}[t]
\centering
    \begin{subfigure}[b]{0.34\linewidth}
        \includegraphics[width=\textwidth]{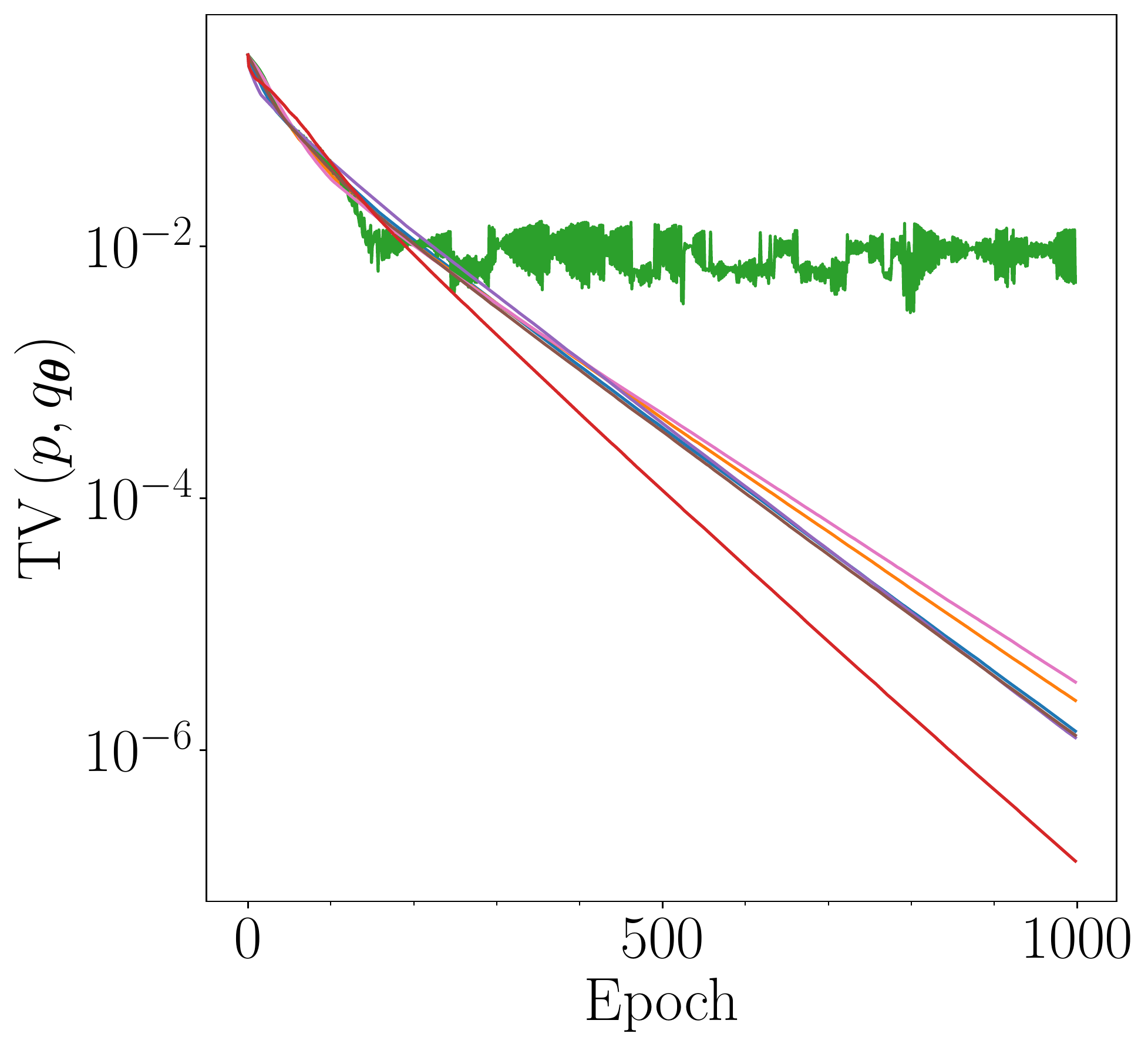}
    \end{subfigure}
    \hspace{8mm}
    \begin{subfigure}[b]{0.34\linewidth}
        \includegraphics[width=\textwidth]{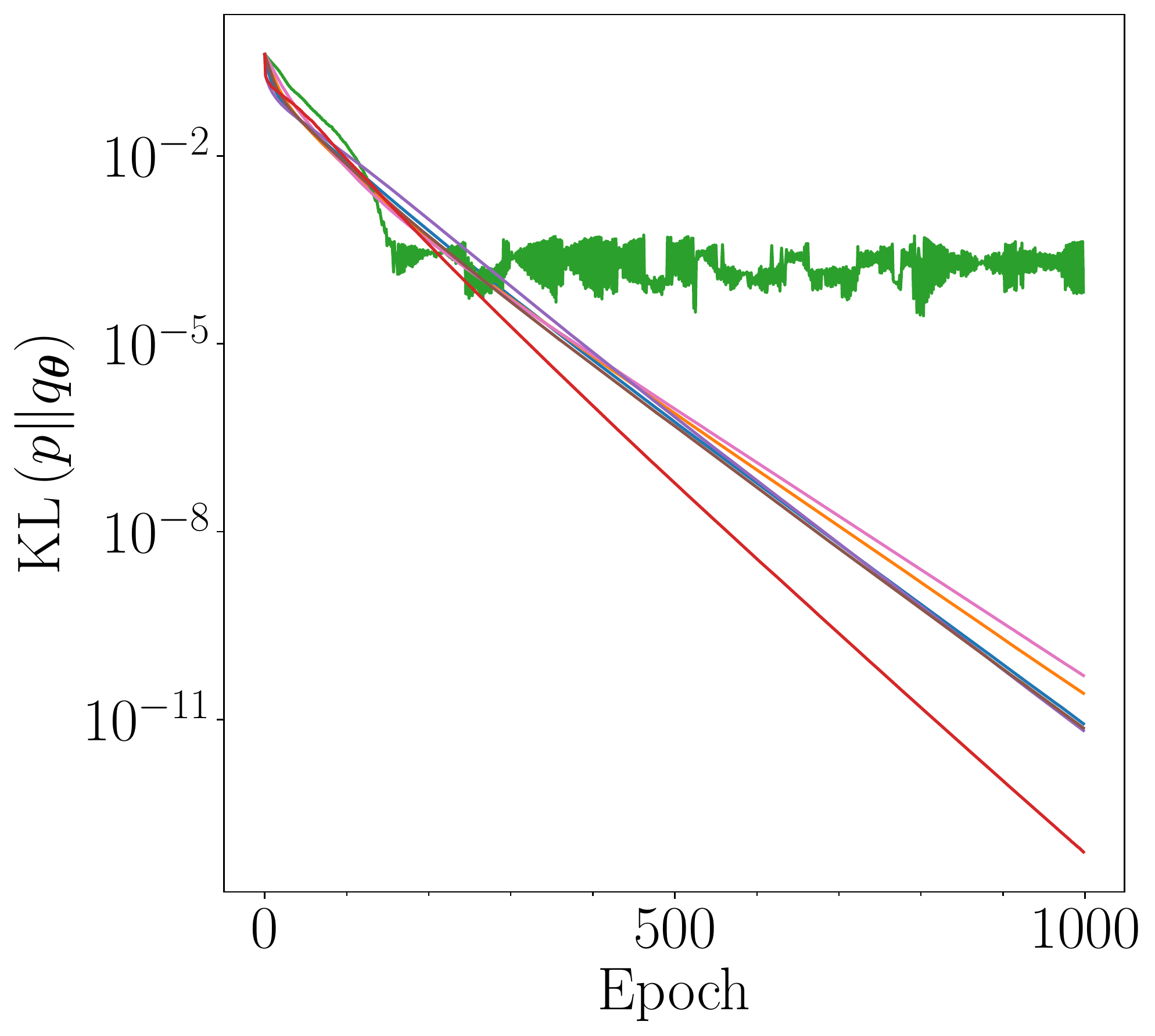}
    \end{subfigure}
    \hspace{8mm}
    \begin{subfigure}[b]{0.14\linewidth}
        \raisebox{0.3\height}{
            \includegraphics[width=\textwidth]{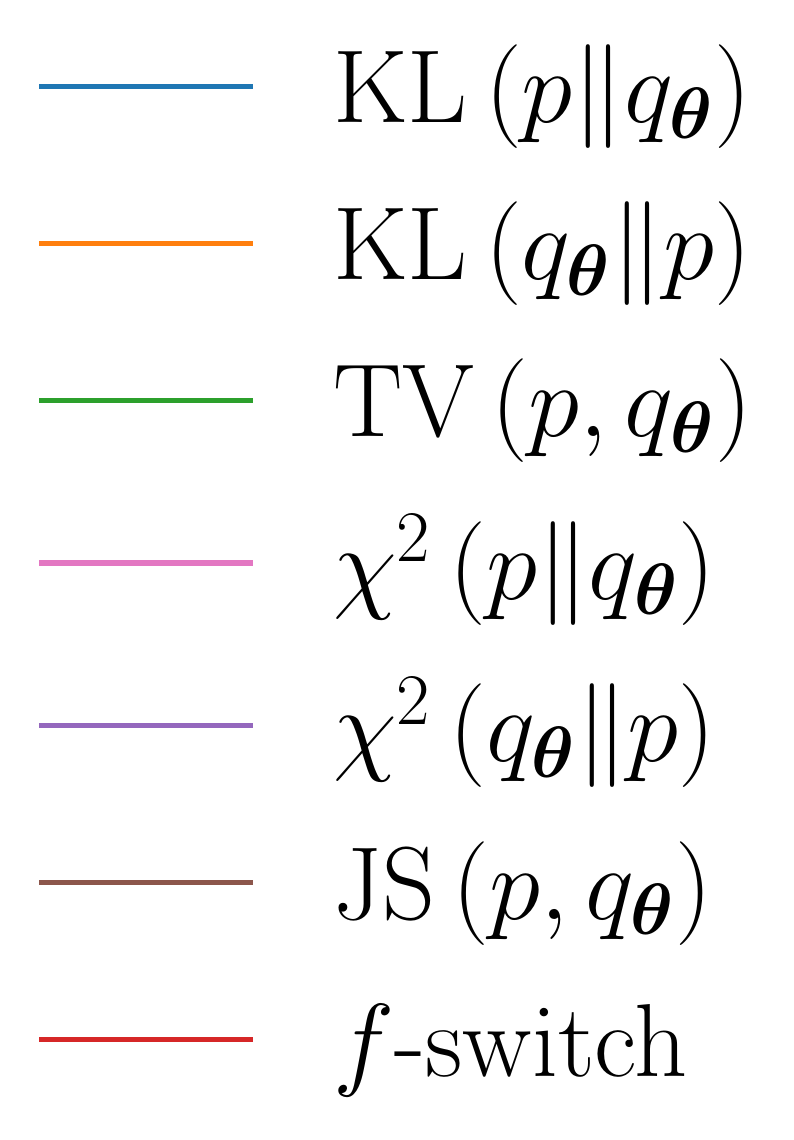}}
    \end{subfigure}
    \caption{Performance of the QCBM trained using several $f$-divergences for 3 qubits in the under-parameterised case U(30,18). The parameters are initialised using the parameters which gave the lowest cost during training in Figure~\ref{fig:U_all_initial_configs}. We show the exact TV (left) and the exact KL (right). }
    \label{fig:U_all_obj}
\end{figure}

The average performance of the heuristic is similar to TV in the exactly and under-parametrised regimes. There are, however, initial parameter configurations within these regimes for which the heuristic significantly outperforms TV. In Figure~\ref{fig:U_all_initial_configs}, we plot the median losses obtained throughout the training of the QCBM in the under-parametrised U(30, 18) regime. The best-performing experiment in this regime is also presented in Figure~\ref{fig:U_all_obj}, alongside all the other $f$-divergences considered in Figure~\ref{fig:generators}. After 200 epochs, the training method that solely uses TV has converged, but all the other divergences, including the heuristic, continue to converge exponentially quickly to smaller losses.
In the under-parameterised regime, the ansatz is not guaranteed to contain the true solution. However, after reaching a KL of $\sim 10^{-3}$, these $f$-divergences traverse similar landscapes. Since the $f$-switch heuristic is shown to reach a KL of $\sim 10^{-5}$, we can assume that all of these $f$-divergences will converge to the global minimum, with the heuristic arriving first.

\begin{figure}[t]
    \begin{subfigure}[b]{0.44\linewidth}
        \includegraphics[width=\textwidth]{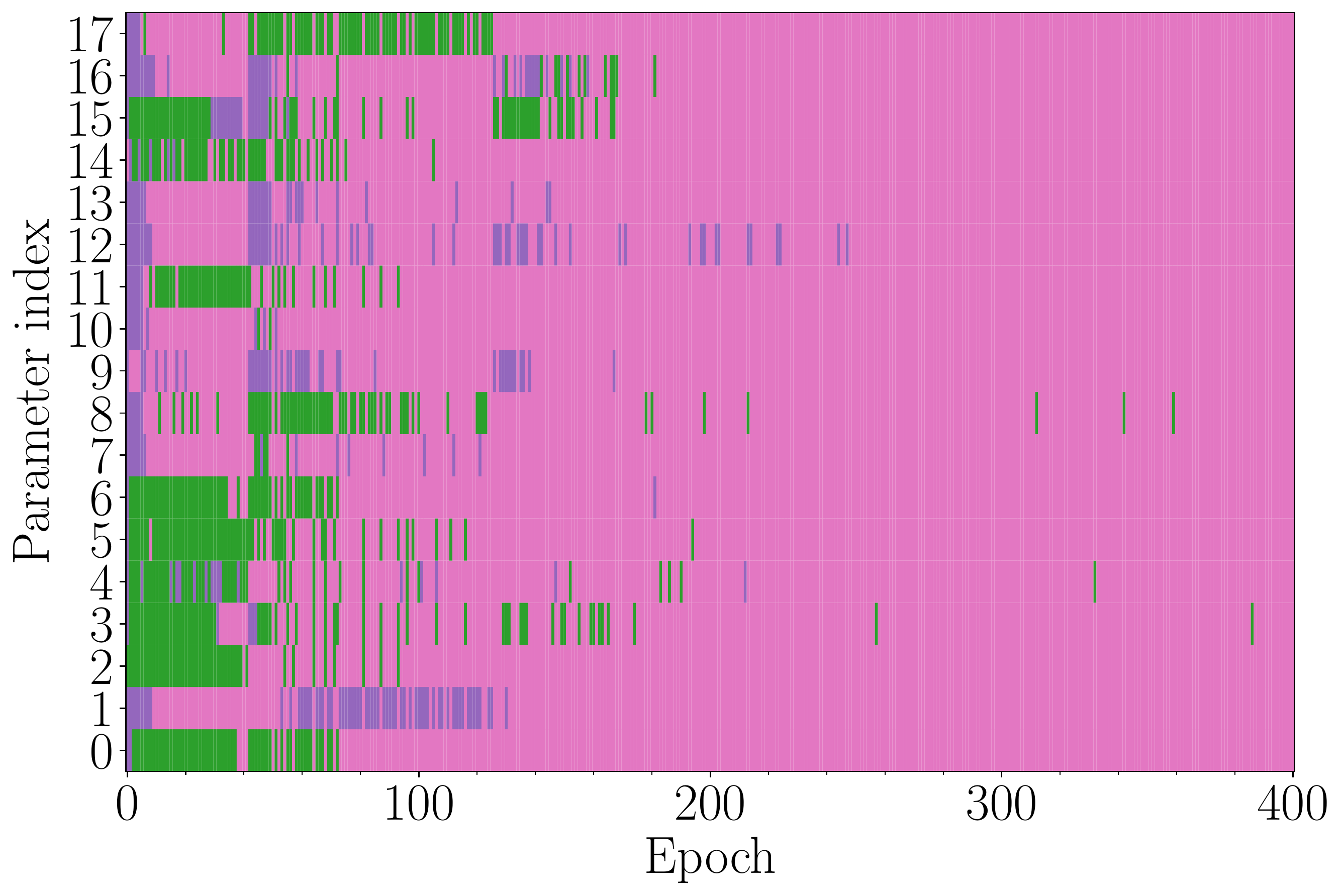}
    \end{subfigure}
    \hspace{8mm}
    \begin{subfigure}[b]{0.14\linewidth}
        \centering
        \raisebox{1.2\height}{
            \includegraphics[width=\textwidth]{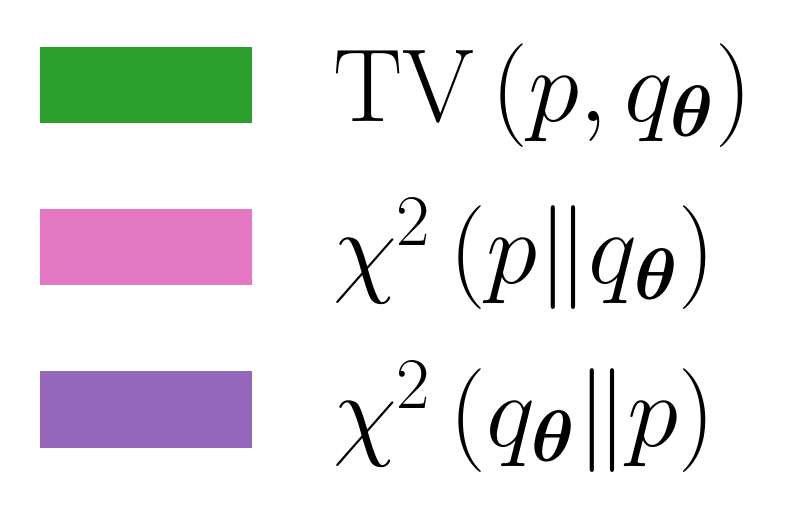}}
    \end{subfigure}
\caption{$f$-divergences chosen throughout the training of the heuristic in Figure~\ref{fig:U_all_obj} in each of the 18~directions in parameter space.}
\label{fig:U_cmap}
\end{figure}

Finally, in Figure \ref{fig:U_cmap}, we illustrate the mechanics of the $f$-switch heuristic. In particular, we plot which $f$-divergence is `activated' for each direction in the parameter space, at each epoch of the training in Figure \ref{fig:U_all_obj}.

We remark that as the number of qubits is increased, the randomly initialised model and the target distributions are expected to be increasingly further apart. The heuristic can pick the divergence that provides the highest initial learning signal. For this reason, we expect the heuristic to become particularly useful as the number of qubits is increased.

\subsection{Local Cost Functions}

\begin{figure}[t]
\captionsetup[subfigure]{justification=centering,skip=0pt}
    \centering
    \subfloat[4 qubits]{\includegraphics[width=.42\linewidth]{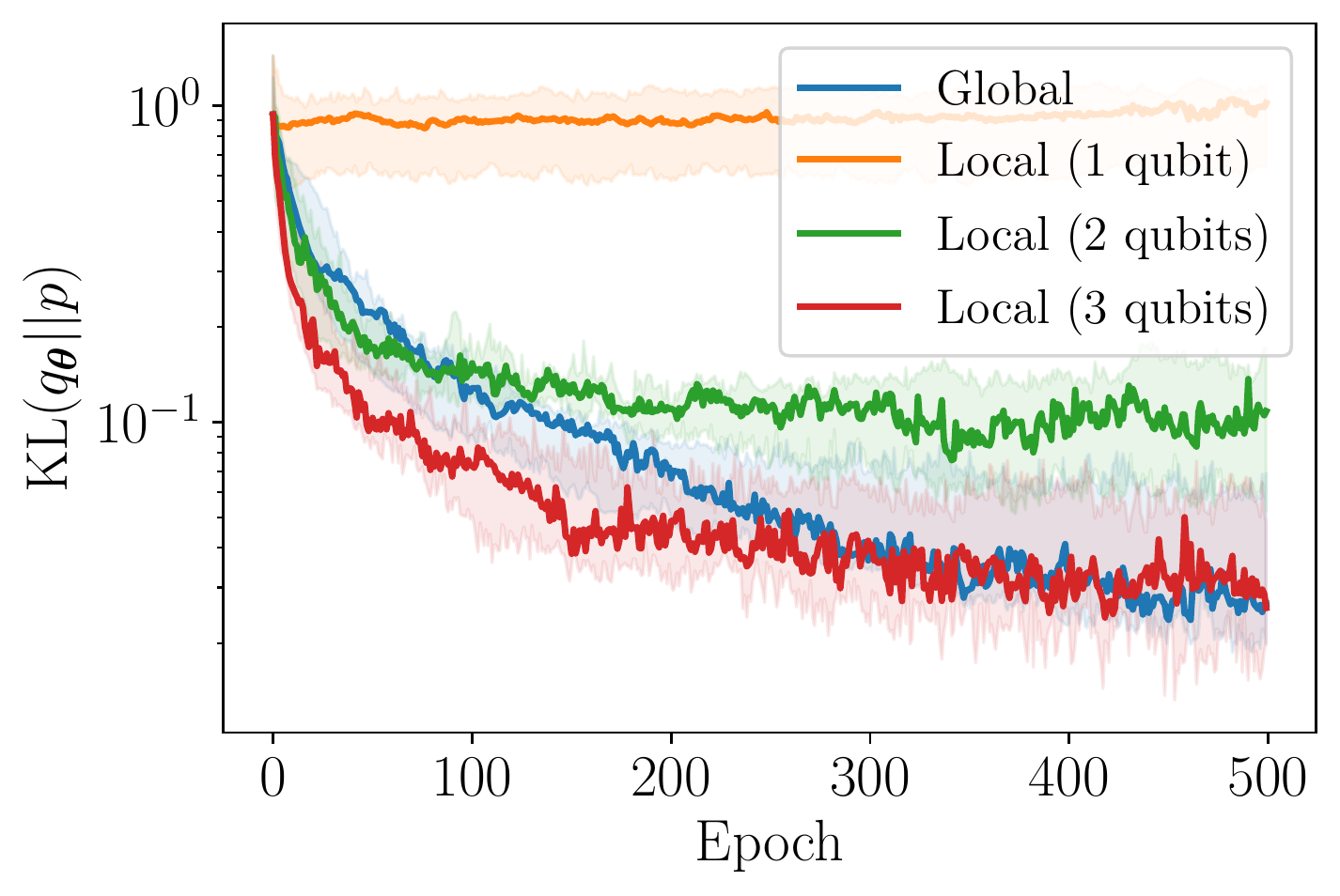}\label{fig:fully_local_1a}}
    \hspace{8mm}
    \subfloat[5 qubits]{\includegraphics[width=.42\linewidth]{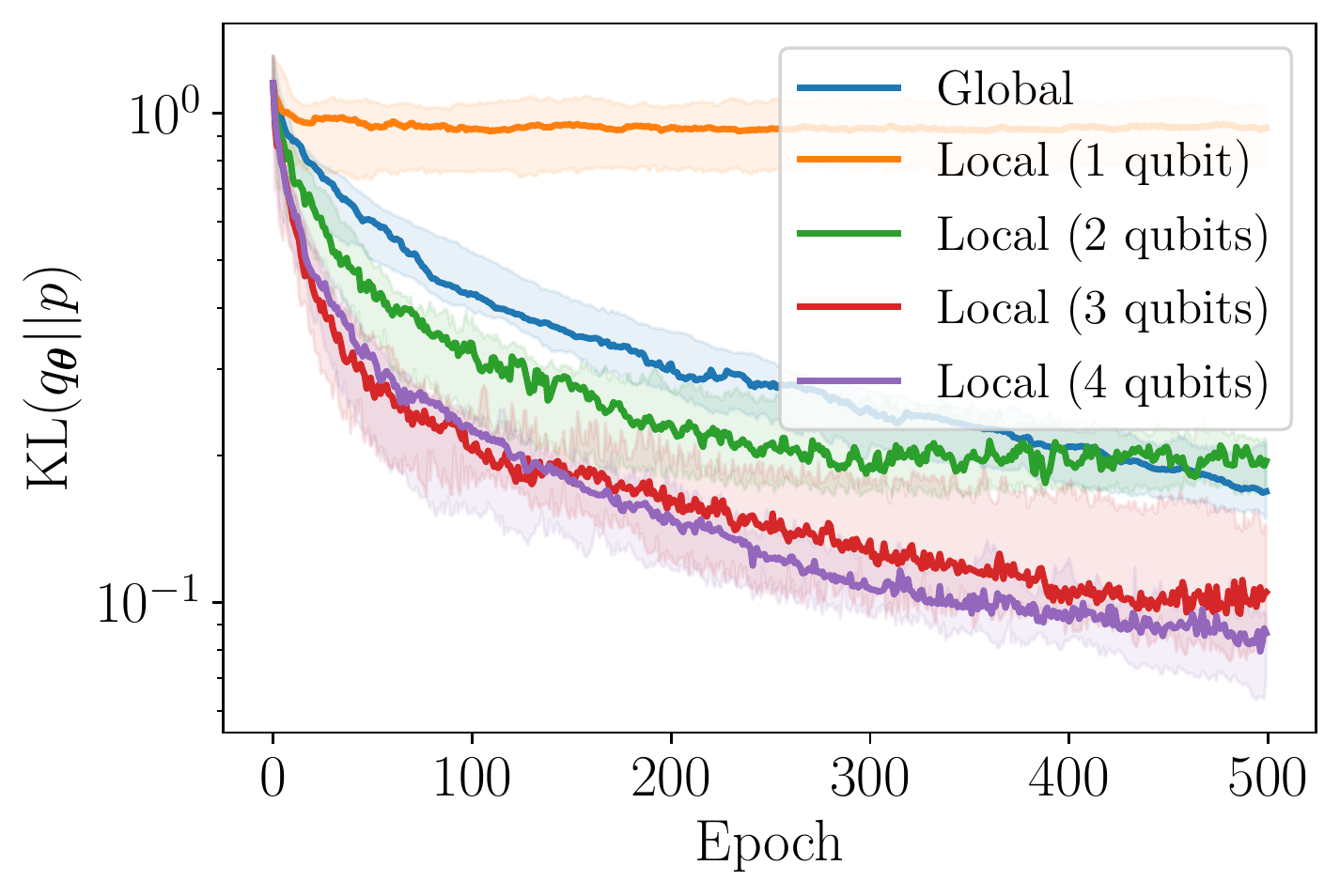}\label{fig:fully_local_1b}}\\
    \vspace{2mm}
    \subfloat[6 qubits \label{fig:fully_local_1c}]{\includegraphics[width=.42\linewidth]{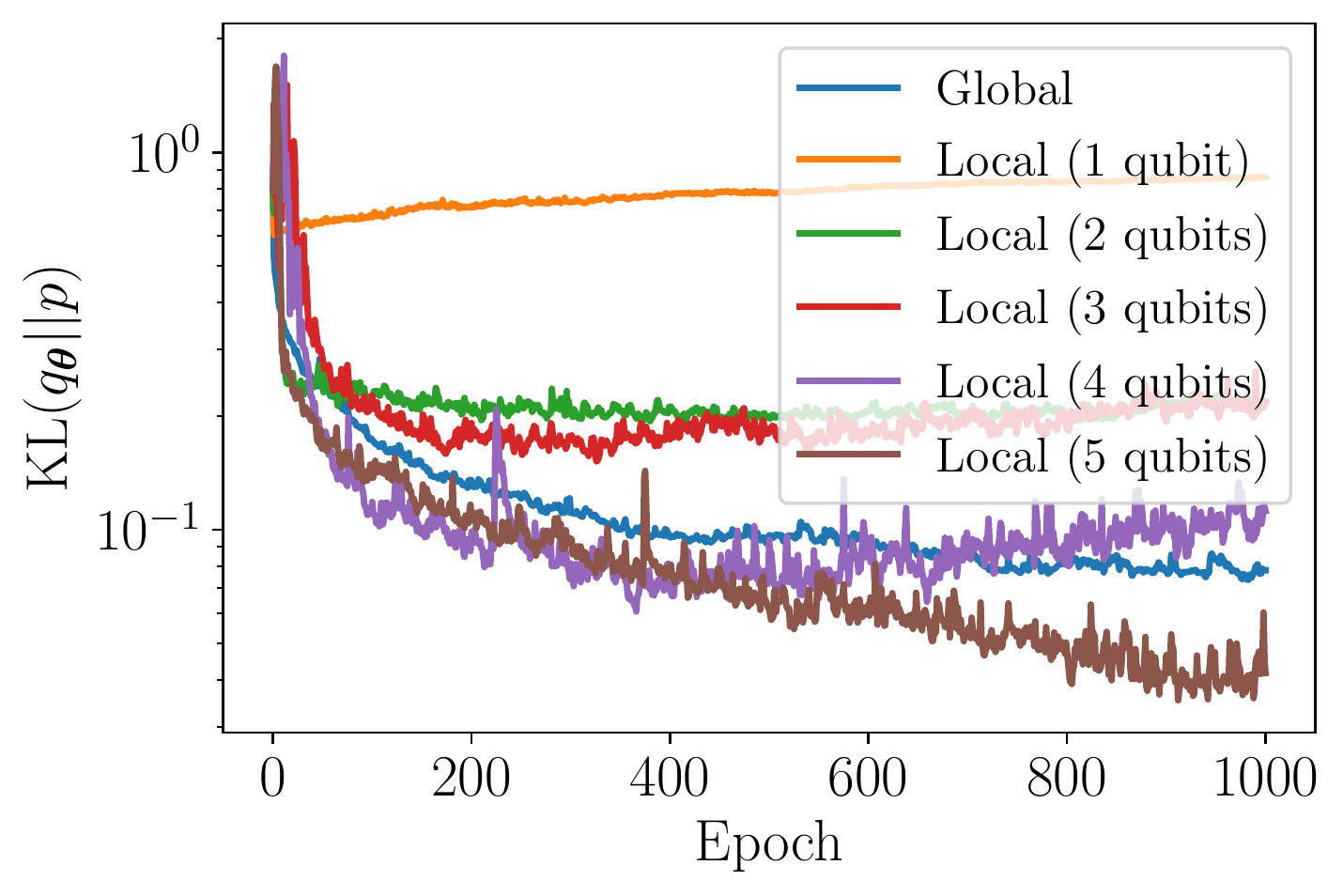}} 
\caption{Training performance of the QCBM using the global and local reverse KL for 4 qubits, 5~qubits, and 6 qubits, for a discretised Gaussian target distribution. For 4 qubits and 5 qubits, we show the bootstrapped  median (solid line), as well as 90\% confidence intervals (shaded). For 6 qubits, we plot an illustrative training example.}
\label{fig:fully_local_comparison}
\end{figure}

We now turn our attention to the heuristic introduced in Section \ref{ssec:local_cost_funcs}, incorporating locality in the cost function, dubbed $f$-local. In this Section, the target distribution is a discretised Gaussian. All classifiers are neural networks with $1$ hidden layer made of $10 k$ ReLU neurons, where $k$ is the locality parameter. The number of layers in the QCBM equals the number of qubits, $D=n$. All expectation values are estimated using $500$ samples.
In Figure \ref{fig:fully_local_comparison}, we plot the training performance of the QCBM using the global cost function and several $k$-local cost functions, for $n=4$, $5$, and $6$ qubit experiments. For $4$ and $5$ qubits, we show the bootstrapped median for the first $500$ training epochs, as well as $90\%$ confidence intervals. For $6$ qubits, we plot an illustrative training example for the first $1000$ training epochs.

Let us make several remarks. Firstly, it would appear that the use of a $k$-local cost function can indeed improve the convergence (rate) of the training procedure, particularly during the initial stages. This improvement is increasingly evident as the number of qubits is increased. As such, this approach could be regarded as a potential strategy for tackling barren plateaus in higher-dimensional problems. However, we leave a thorough study of this phenomenon to future work.

Secondly, it is clear that the use of any $k$-local cost function will eventually prohibit convergence to the true target distribution. As discussed in Section~\ref{ssec:local_cost_funcs}, the $k$-local cost function is minimised whenever the $k$-marginal distributions of the target and the model coincide, which does not necessarily imply that their joint distributions are equal. The smaller the value of $k$, the greater the possible disparity between two distributions whose $k$-marginals coincide. This is clearly visualised in Figure \ref{fig:fully_local_comparison}: as the value of $k$ decreases, the asymptotic reverse KL achieved during training with the $k$-local cost function plateaus at increasingly larger values. 

As remarked previously, this suggests that an optimal training strategy may be to start the training procedure with a small value of $k$, before iteratively increasing the value of $k$ as training proceeds. For example, let us consider the $5$ qubit experiment in Figure \ref{fig:fully_local_1b}. Initially, the $3$-local cost function (red) appears to yields the greatest convergence rate. After approximately 150 epochs, the $4$-local cost function (purple) now seems to be favourable. Asymptotically, one can imagine that the global cost function (blue) will be preferable. One observes similar behaviour in the $6$ qubit experiment in Figure \ref{fig:fully_local_1c}. 

In practice, of course, it is not possible to compute the reverse KL directly, and thus another tractable metric is required in order to determine the optimal moment for switching between the $k$-local cost functions. Alternatively, one can simply increase the locality of the cost function after a set number of epochs.

\section{Estimation of \texorpdfstring{$f$}{f}-divergences on Fault-Tolerant Quantum Computers} 
\label{sec:fault_tolerant_est}

The above discussion is purely heuristic in nature and suitable for near-term quantum computers, but we can also address $f$-divergences from the other end of the spectrum; using fault-tolerant devices. In particular, we can leverage a recent line of study into quantum property testing of distributions. The key question here is whether or not a particular probability distribution has a certain property.

The work of~\cite{bravyi_quantum_2011} was one of the first to provide such an answer, demonstrating a quadratic speedup for determining whether two distributions over $[n]$ were close or $\epsilon$-far in TV. These quantum algorithms typically work in the \emph{oracle} model, and we measure run time relative to the number of queries to such an oracle (query complexity). In the classical case, we define oracle access to a distribution over $[n]$, $p=\{p_i\}_{i=1}^n$ as $O_p:[S] \rightarrow [n], S \in \mathbb{N}$. The oracle is a mechanism to translate a uniform distribution over $[S]$ to the true distribution over $[n]$. In the quantum case, such an oracle is replaced by a unitary operator, $\hat{O}_p$ acting on a state encoding $s\in [S]$, along with an ancillary register to ensure reversibility and defined as: $\hat{O}_p\ket{s}\ket{0} = \ket{s}\ket{O_p(s)} \forall s \in [S]$.

We begin our discussion with the TV. The authors of~\cite{bravyi_quantum_2011} produced a quantum property testing algorithm for the TV via an algorithm which actually \emph{estimates} the TV quadratically faster. The analysis in~\cite{bravyi_quantum_2011} resulted in an algorithm to estimate the TV up to additive error $\varepsilon$, with probability of success of $1-\delta$, using $O({\sqrt{n}}/{\epsilon^8\delta^5})$ samples. This was later improved by~\cite{montanaro_quantum_2015} to the following

\begin{theorem}[Section 4, Montanaro~\cite{montanaro_quantum_2015}]
\label{thm:tvd_quantum_algorithm}
    Assume $p, q$ are two distributions on $[n]$. Then there is a quantum algorithm that approximates $TV(p, q)$ up to an additive error $\epsilon >0$, with probability of success $1-\delta$, using ${O}({\sqrt{n}}{\epsilon^{-3/2}}/\log\left(1/\delta\right))$ quantum queries.
\end{theorem}

These ideas were extended in~\cite{li_quantum_2019} to also give an algorithm for computing the (forward) KL quadratically faster than possibly classically (and also computing certain entropies of distributions). Due to the existence of the ratio ${p_i}/{q_i}$ in the expression for the KL, we must make a further assumption, which was not necessary in the case of the TV distance in Theorem~\ref{thm:tvd_quantum_algorithm}. This assumption will also be necessary when considering many of the other divergences in Table~\ref{t:f_divs}. In particular, we must assume the two distributions are such that: ${p_i}/{q_i} \leq g(n),~\forall i \in [n]$, for some $g: \mathbb{N} \rightarrow \mathbb{R}^+$.\ (This assumption is appropriate when one defines the KL in terms of the generator $f$ and the ratio $r = p/q$. Conversely, when one defines the KL in terms of the conjugate $f^{*}$ and the ratio $r=q/p$, then the appropriate assumption would instead be that $q_i/p_i\leq g(n)$, $\forall i \in[n]$.) This assumption is also necessary in the classical case. With this, we then have

\vspace{10mm}
\begin{theorem}[Theorem 4.1, Li and Wu~\cite{li_quantum_2019}]
\label{thm:kl_div_quantum_algorithm}
    Assume $p, q$ are two distributions on $[n]$ satisfying ${p_i}/{q_i} \leq g(n),~\forall i \in [n]$ for some $a: \mathbb{N} \rightarrow \mathbb{R}^+$. Then there is a quantum algorithm that approximates $KL(p \| q)$ within an additive error $\epsilon >0$ with probability of success at least $2/3$ using $\widetilde{O}({\sqrt{n}}/{\epsilon^2})$ quantum queries to $p$ and $\widetilde{O}({\sqrt{n}\hspace{.5mm}g(n)}/{\epsilon^2})$ quantum queries to $q$. (The notation $\widetilde{O}(\cdot)$ ignores factors that are polynomial in $\log n$ and $\log 1/\varepsilon$.)
\end{theorem}

These results cover two of the $f$-divergences we use above (see Table \ref{t:f_divs}). In particular, the latter algorithm provide a quantum speedup since it is known that one requires $\Omega\left(n/\log(n)\right), \Omega\left(ng(n)/\log(n)\right)$ \emph{classical} queries to $p$ and $q$ respectively to estimate the KL~\cite{han_minimax_2016}. On the other hand, we get a speedup for the former algorithm since it is known one requires $\Theta(n^{2/3} \varepsilon^{-4/3})$~\cite{chan2013optimal} queries to test if two distributions are near or far in TV classically, which is an easier problem than estimating the metric directly.

The key idea behind both of these algorithms is to use a subroutine known as \emph{quantum probability estimation} or \emph{quantum counting}, which is adapted from \emph{quantum amplitude estimation}. This provides a quadratic speedup in producing estimates $\tilde{p}_i, \tilde{q}_i$, of probabilities $p_i,q_i$ from the distributions $p,q$, which are specified via a quantum oracle. Once the estimates of $\tilde{p}_i,\tilde{q}_i$ have been produced via the quantum subroutine, both of the above algorithms reduce to simple classical post-processing. This post processing involves constructing a random variable, $y$, whose expectation value gives exactly the divergence we require. For TV and KL estimation, this random variable is given by
\begin{eqnarray}
    y^{\text{TV}}_i &:=& \frac{\left|p_i - q_i\right|}{p_i + q_i}, \\
    y^{\text{KL}}_i &:=& \log\frac{p_i}{q_i} = \log p_i - \log q_i .
\label{eqn:tvd_kl_random_variable}
\end{eqnarray}
By sampling this random variable according to another distribution $r := (r_i)_{i=1}^n$ (to be defined below), the quantity of interest is exactly given as an expectation value, namely
\begin{eqnarray}
    \sum\limits_i r^{\text{TV}}_i y^{\text{TV}}_i  &=& \mathbb{E}[y^{\text{TV}}] = \text{TV}(p, q), \\
    \sum\limits_i r^{\text{KL}}_i y^{\text{KL}}_i  &=& \mathbb{E}[y^{\text{KL}}] = \text{KL}(p \| q) .
\end{eqnarray}
One can check~\cite{bravyi_quantum_2011, li_quantum_2019} that the suitable random variables are given by
\begin{eqnarray}
    r^{\text{TV}}_i &=& \frac{1}{2}\left(p_i + q_i\right), \\
    r^{\text{KL}}_i &=& q_i .
\label{eqn:tvd_kl_mod_distributions}
\end{eqnarray}
Due to the probabilistic nature of quantum mechanics, one cannot obtain the \emph{exact} values of the probabilities required to compute these expectation values. We must settle instead for \emph{approximations} of $p, q$, namely $\tilde{p}, \tilde{q}$. These estimates are achieved using the \emph{quantum approximate counting} lemma, which is an application of quantum amplitude estimation~\cite{brassard_quantum_2002}. The work in~\cite{li_quantum_2019} considered two versions of this algorithm, called \textsf{EstAmp} and \textsf{EstAmp'}. The only difference between these two algorithms is the behavior when one of the probabilities, $q_i$, is sufficiently close to zero. This is problematic in the case of the KL estimation (and indeed entropy estimation) in~\cite{li_quantum_2019} since the relevant quantities diverge as $q_i\rightarrow 0$. The same is true in our case, as $q_i^{-1}$ appears in many $f$-divergences.

\begin{theorem}[Theorem 13, Brassard et al.~\cite{brassard_quantum_2002} and Theorem 2.3, Li and Wu~\cite{li_quantum_2019}]
For any $k, M \in \mathbb{N}$, there is a quantum algorithm (named \textsf{EstAmp}) with $M$ queries to a boolean function, $\chi:[S] \rightarrow \{0, 1\}$ that outputs $\tilde{a}=\sin^2(\frac{l\pi}{M})$ for some $l \in \{0, \dots, M-1\}$ such that
\begin{equation}
    \text{Pr}\left[\tilde{a} = \sin^2\left(\frac{l\pi}{M}\right)\right] = \frac{\sin^2(M\Delta\pi)}{M^2\sin^2(\Delta \pi)} \leq \frac{1}{(2M\Delta)^2} ,
\end{equation}
where $\Delta = |\omega - l/M|$. This promises $|\tilde{a} - a| \leq 2\pi k \frac{\sqrt{a(1-a)}}{M} + k^2\frac{\pi^2}{M^2}$ with probability at least $8/\pi^2$ for $k=1$ and with probability greater than $1-\frac{1}{2(k=1)}$ for $k \geq 2$. If $a=0$ then $\tilde{a}=0$. 
\end{theorem}

The modified algorithm (\textsf{EstAmp}') outputs $\sin^2(\frac{\pi}{2M})$ when \textsf{EstAmp} outputs $0$, and outputs the same as \textsf{EstAmp} otherwise. Now that we have a mechanism for estimating the probabilities, we need a final ingredient, which is the generic speedup of Monte Carlo methods from~\cite{montanaro_quantum_2015}

\begin{theorem}[Theorem 5, Montanaro~\cite{montanaro_quantum_2015}]
\label{thm:montanaro_montecarlo_speedup}
Let $\mathcal{A}$ be a quantum algorithm with output $X$ such that $\text{Var}[X] \leq \sigma^2$. Then for $\epsilon$ where $0 < \epsilon < 4\sigma$, by using $O((\sigma/\epsilon) \log^{3/2}(\sigma/\epsilon)\log\log(\sigma/\epsilon))$ executions of $\mathcal{A}$ and $\mathcal{A}^{-1}$, Algorithm $3$ in~\cite{montanaro_quantum_2015} outputs an estimate $\tilde{\mathbb{E}}[X]$ of $\mathbb{E}[X]$ such that
\begin{equation} 
\label{eqn:montanaro_montecarlo_speedup}
    \text{Pr}\left[|\tilde{\mathbb{E}}[X] - \mathbb{E}[X]| \geq \epsilon\right] \leq 1/5 .
\end{equation}
\end{theorem}

Using these results, we now extend Theorems~\ref{thm:tvd_quantum_algorithm}, \ref{thm:kl_div_quantum_algorithm} to cover another $f$-divergence in Table~\ref{t:f_divs}: the forward Pearson divergence, $\chi^2(p\|q)$. The conjugate of the generator for this divergence is given by $f^{*}(r) = \frac{1}{2}(r-1)^2$ or, equivalently, $f^{*}(r) = \frac{1}{2}(r^2-1)$. The equivalence of these two generators is straightforward to demonstrate. In particular, we have
\begin{equation}
    \mathbb{E}_{p}\left[\left(\frac{q_i}{p_i}-1\right)^2\right] = \sum_{i} p_i\left(\frac{q_i}{p_i}\right)^2 - 2\sum_{i} p_i\frac{q_i}{p_i} + \sum_{i}p_i = \sum_{\boldsymbol{x}} p_i\left(\frac{q_i}{p_i}\right)^2 - 1
    =  \mathbb{E}_{p}\left[\left(\frac{q_i}{p_i}\right)^2-1\right] .
\end{equation}
In fact, in what follows, we make use of the following representation
\begin{equation}
   {\chi}^2(p \| q) := \frac{1}{2}\sum_{i} p_i \left[\left(\frac{q_i}{p_i}\right)^2 - 1\right] = \sum_i q_i \left(\frac{q_i}{p_i}-1\right): = \sum_{i} r_i^{FP} y_i^{FP} ,
\end{equation} 
where we have identified $r_i^{FP} = q_i$ and $y_i^{FP} = \frac{1}{2}(\frac{q_i}{p_i}-1)$. Using this representation, we develop the following algorithm for estimating the forward Pearson divergence.

\SetKwInput{KwSet}{Set}
\SetKwInput{KwLet}{Let}
\SetKwInput{KwFor}{For}
\begin{algorithm}
    \KwSet{ $l = \Omega\left((\sigma/\epsilon) \log^{3/2}(\sigma/\epsilon)\log\log(\sigma/\epsilon)\right), \sigma := g(n)^2\left[1+ \frac{\exp\left(-\frac{\epsilon^2}{2n}\right)}{g(n)^2}\right]$.}
    \KwSet{The following subroutine to be the algorithm $\mathcal{A}$:}
    \Begin{
    Sample an index, $i \in [n]$, according to p.
    Use the procedure \textsf{EstAmp}' with $2^{\lceil\log_{2}(\sqrt{n}g(n)/\varepsilon)\rceil}$ and $2^{\lceil\log_{2}(\sqrt{n}g(n)^2/\varepsilon)\rceil}$ queries to $q$ and $p$, respectively. Obtain estimates of $\tilde{p}_i, \tilde{q}_i$.
    Output $\tilde{y}_i^{\text{FP}} = \frac{1}{2}\left(\frac{\tilde{q}_i}{\tilde{p}_i} - 1\right)$.
    }
    Use $\mathcal{A}$ for $l$ times in Theorem~\ref{thm:montanaro_montecarlo_speedup} to output an estimate, $\tilde{\chi}^2(p \| q)$ for $\chi^2(p \| q)$.
    \caption{Estimate the forward Pearson divergence of $p= (p_i)_{i=1}^n$ and $(q_i)_{i=1}^n$ on $[n]$.}
\label{alg:forward_pearson_alt_2}
\end{algorithm}

The query complexity of this Algorithm is contained in the following theorem. We defer the proof of this result, which is largely a technical extension of the proof(s) in~\cite{li_quantum_2019}, to Appendix~\ref{app:proof_of_pearson}.

\begin{theorem} \label{thm:pearson_algorithm}
    Assume $p, q$ are two distributions on $[n]$ satisfying ${q_i}/{p_i} \leq g(n),~\forall i \in [n]$ for some $a: \mathbb{N} \rightarrow \mathbb{R}^+$. Then there is a quantum algorithm that approximates $\mathcal{X}^2(p \| q)$ within an additive error $\epsilon >0$ with probability of success at least $2/3$ using $ \tilde{{O}}(\sqrt{n}g(n)/\varepsilon^2)$ quantum queries to $q$ and $ \tilde{{O}}(\sqrt{n}g(n)^2/\varepsilon^2)$ quantum queries to $p$.
\end{theorem}

\section{Discussion}
\label{sec:discussion}

Each $f$-divergence, with its unique operational meaning, finds application in information theory, statistics, and machine learning. In this paper, we showed that a generative model called quantum circuit Born machine can be trained by efficiently minimising \emph{any} $f$-divergence. The key observation is that a probabilistic classifier can be trained adversarially to provide an approximation to such divergences.

Building on this, we developed heuristics aimed at improving convergence of the generative training. 
The first heuristic, \emph{f-switch}, lets each parameter minimise a different $f$-divergence. Numerical results with an ideal exact classifier show that this heuristic can converge faster and to better minima than when using a single $f$-divergence. However, in a more realistic setting where the classifier is trained adversarially, $f$-switch yields results similar to those obtained by minimising a single $f$-divergence. 

The second training heuristic, \emph{f-local}, consists of using a single $f$-divergence approximated by local cost functions. Numerical results show that, as the number of qubits increases, this strategy yield improved convergence of the generative training than when using a global cost function. To the best of our knowledge this is the first proposal of cost functions for generative modelling that can interpolate between trainability and accuracy. Extensive numerical simulations will be needed to confirm whether $f$-local can alleviate the barren plateau problem in generative modelling.

Interestingly, our local cost functions approximate the $f$-divergence using an ensemble of local binary classifiers. If the target probability distribution is known to have a particular conditional independence structure (e.g., it is defined by a Bayesian network or a Hidden Markov model), this information could be used to inform the choice of local classifiers. 

One interesting research direction is to adapt the above heuristics to work with other families of distance measures. Of particular interest, integral probability metrics (IPMs) include the maximum mean discrepancy, the Dudley metric and the Wasserstein distance. While $f$-divergences are defined in terms of probability ratios, IPMs are defined in terms of probability differences. However, it is know that under suitable constraints margin-based classifiers yield estimators for IPMs~\cite{sriperumbudur2009integral}. This suggests that an extension of our heuristics to IPMs could be possible.

In this work, we also discussed the possibility of estimating certain $f$-divergences on a fault-tolerant quantum computer, therefore avoiding the use of classifiers. Previously published work has proven quadratic quantum speedups for the estimation of total variation~\cite{bravyi_quantum_2011,montanaro_quantum_2015} and forward Kullback-Leibler (KL) of type I~\cite{li_quantum_2019}. Using these algorithms a quadratic speedup is achievable for the reverse KL of type I, and thus for the symmetric KL of type I (also known as Jeffrey divergence). It is plausible that with some refinements these algorithms can provide a quadratic speedup for the KL of type II as well. 

We contributed to this topic with an algorithm for estimating Pearson $\chi^2$ divergences and by providing its query complexity. Interestingly, high-order Pearson divergences (also known as Vajda divergences) can be used to approximate any other $f$-divergence via Taylor expansion~\cite{nielsen2014chisquare}. Generalising our quantum algorithm to Vajda divergences would therefore provide a way to estimate all other $f$-divergences on a fault-tolerant quantum computer.

\section*{Acknowledgments}
This research was funded by Cambridge Quantum Computing. M.B. would like to thank Mattia Fiorentini for helpful conversations. 

\section*{Author Contributions}
C.L. devised and implemented the $f$-switch heuristic. L.S. devised and implemented the $f$-local heuristic. B.C. and L.S. devised and analysed the fault-tolerant algorithm. C.L. and L.S. designed the figures. M.B. and B.C. supervised the work. All authors analysed the results and contributed to the final manuscript.

\section*{Data Availability}
Data used to generate the above ﬁgures are available upon request from the authors.

\section*{Competing Interests}
The authors declare no conflict of interest.

\appendix

\section{Proof of Theorem~ \ref{thm:pearson_algorithm}} 
\label{app:proof_of_pearson}

In this Appendix, we provide a proof of Theorem~\ref{thm:pearson_algorithm}. For completeness, we first repeat the theorem here.

\begin{customtheorem}{5} 
\label{thm:pearson_algorithm_app}
    Assume $p, q$ are two distributions on $[n]$ satisfying ${q_i}/{p_i} \leq g(n),~\forall i \in [n]$ for some $a: \mathbb{N} \rightarrow \mathbb{R}^+$. Then there is a quantum algorithm that approximates $\mathcal{X}^2(p \| q)$ within an additive error $\epsilon >0$ with probability of success at least $2/3$ using $ \tilde{\mathcal{O}}(\sqrt{n}g(n)/\varepsilon^2)$ quantum queries to $q$ and $ \tilde{\mathcal{O}}(\sqrt{n}g(n)^2/\varepsilon^2)$ quantum queries to $p$.
\end{customtheorem}

\begin{proof}
We prove this theorem in two parts, following closely the approach in~\cite{li_quantum_2019}. We are first required to show that the expectation of the output of the sub-routine $\mathcal{A}$, namely $\tilde{E} = \sum_{i\in[n]} q_i (\tilde{q}_i/\tilde{p}_i - 1)$ is sufficiently close to $E = \sum_{i\in[n]} q_i (q_i/p_i - 1)$. We begin by observing the following inequality. Let $x,y>0$. In addition, suppose there exists $0<K<\infty$ such that $y\leq K$. Then
\begin{align}
    \left|y-x\right| \leq K \left|\frac{y-x}{y}\right| =  K \left|\frac{{y}/{x}-1}{{y}/{x}}\right| \leq K \left|\log \left(\frac{y}{x}\right)\right| = K \left|\log(y) - \log(x)\right| .
\end{align}
where we have used the elementary inequality: $(z-1)/z\leq \log(z)$. We then have, using the linearity of expectation,
\begin{align}
    |E - \tilde{E}| &\leq \frac{1}{2} \sum_{i\in[n]} q_i \mathbb{E}\left[ \left|\left(\frac{q_i}{p_i}\right) - \left(\frac{\tilde{q}_i}{\tilde{p}_i}\right)\right|\right] \\
    &\leq \frac{1}{2}g(n) \sum_{i\in[n]} q_i \mathbb{E}\left[\left|\log\left(\frac{q_i}{p_i}\right) - \log \left(\frac{\tilde{q}_i}{\tilde{p}_i}\right)\right|\right] .
\end{align}

The remainder of the proof follows~\cite{li_quantum_2019}, with the roles of $p$ and $q$ now reversed, and with an additional factor of $g(n)$. In particular, using elementary properties of the logarithm, we have
\begin{align}
    |E - \tilde{E}| &\leq \frac{1}{2}g(n) \sum_{i\in[n]} q_i \mathbb{E}\left[\left|\log q_i - \log \tilde{q}_i\right|\right] + \frac{1}{2}g(n)\sum_{i\in[n]} q_i \mathbb{E}\left[\left|\log p_i - \log \tilde{p}_i\right|\right] \\
    &\leq \frac{1}{2}g(n) \sum_{i\in[n]} q_i \mathbb{E}\left[\left|\log q_i - \log \tilde{q}_i\right|\right] + \frac{1}{2}g(n)^2 \sum_{i\in[n]} p_i \mathbb{E}\left[\left|\log p_i - \log \tilde{p}_i\right|\right] ,
\label{eq:exp_bound}
\end{align}
where in the second line we have used the assumption that $q_i/p_i\leq g(n)$ for all $i\in[n]$. By (IV.5) and (IV.6) in~\cite[Section IV]{li_quantum_2019}, $2^{\lceil\log_{2}(\sqrt{n}g(n)/\varepsilon)\rceil}$ queries to $q$ and $2^{\lceil\log_{2}(\sqrt{n}g(n)^2/\varepsilon)\rceil}$ queries to $p$ yield
\begin{align}
    \sum_{i\in[n]} q_i \mathbb{E}\left[\left|\log q_i - \log \tilde{q}_i\right|\right] &= \mathcal{O}\left(\frac{\varepsilon}{g(n)}\right), \\
    \sum_{i\in[n]} p_i \mathbb{E}\left[\left|\log p_i - \log \tilde{p}_i\right|\right] &= \mathcal{O}\left(\frac{\varepsilon}{g(n)^2}\right) .
\end{align}

Substituting these bounds into Equation~\eqref{eq:exp_bound}, and re-scaling Algorithm~\eqref{alg:forward_pearson_alt_2} by a large enough constant, we obtain $|E - \tilde{E}|\leq \frac{\varepsilon}{2}$. We are now required to bound the variance of this random variable. The variance is at most
\begin{align}
   \frac{1}{4}\sum_{i \in [n]} q_i \left(\frac{\tilde{q}_i}{\tilde{p}_i} - 1\right)^2 = \frac{1}{4}\sum_{i \in [n]: \tilde{q}_i\leq \tilde{p}_i} q_i \left(\frac{\tilde{q}_i}{\tilde{p}_i} - 1\right)^2 + \frac{1}{4}\sum_{i \in [n]: \tilde{q}_i> \tilde{p}_i} q_i \left(\frac{\tilde{q}_i}{\tilde{p}_i} - 1\right)^2 .
\label{eq:variance}
\end{align}

We first turn our attention to the first term. Recall that \textsf{EstAmp}' outputs $\tilde{q}_i$ such that $\tilde{q}_i \geq \sin^2(\pi / 2^{\lceil\log_{2}(\sqrt{n}g(n)/\varepsilon)\rceil+1})\geq{\epsilon^2}/{(4ng(n)^2)}$ for any $i$. It follows that ${\tilde{q}_i}/{\tilde{p}_i}\geq \tilde{q}_i\geq \varepsilon^2/(4ng(n)^2)$, and thus $\exp(-2{\tilde{q}_i}/{\tilde{p}_i})\leq \exp(- \varepsilon^2/(2ng(n)^2))$. We thus have, using also the fact that $(x-1)^2 < \exp(-2x)$ for $x>-1$, that
\begin{align}
    \frac{1}{4}\sum_{i: \tilde{q}_i < \tilde{p}_i} q_i \left(\frac{\tilde{q}_i}{\tilde{p}_i} - 1\right)^2  &\leq \frac{1}{4}\sum_{i: \tilde{q}_i < \tilde{p}_i} q_i \exp\left(-2\frac{\tilde{q}_i}{\tilde{p}_i}\right)  
\label{eqn:var_first_term_proof_line1}\\
    &\leq \frac{1}{4}\sum_{i: \tilde{q}_i < \tilde{p}_i} q_i  \exp\left(-\frac{\epsilon^2}{2ng(n)^2}\right)  \leq \exp\left(-\frac{\epsilon^2}{2ng(n)^2}\right) .
\label{eqn:var_first_term_proof_line2}
\end{align}

Meanwhile, for the second term, using the fact that $(\tilde{q}_i/\tilde{p}_i-1)^2 \leq [\tilde{q}_i/\tilde{p}_i]^2$ since, in this summation, $\tilde{q}_i/\tilde{p}_i\geq 1\geq 1/2$, we obtain
\begin{align}
    \frac{1}{4}\sum_{i: \tilde{q}_i \geq \tilde{p}_i} q_i \left(\frac{\tilde{q}_i}{\tilde{p}_i} - 1\right)^2 \leq \frac{1}{4}\sum_{i: \tilde{q}_i \geq \tilde{p}_i} q_i \left[\frac{\tilde{q}_i}{\tilde{p}_i}\right]^2
    \leq \frac{1}{4}\sum_{i: \tilde{q}_i \geq \tilde{p}_i} q_i g(n)^2  \leq g(n)^2 .
\label{eqn:var_second_term_proof_line2}
\end{align}

Substituting Equations~\eqref{eqn:var_first_term_proof_line2} and~\eqref{eqn:var_second_term_proof_line2} into Equation~\eqref{eq:variance}, we see that the variance of the random variable is at most
\begin{align}
    g(n)^2 +  \exp\left(-\frac{\epsilon^2}{2ng(n)^2}\right) = O\left(g(n)^2\left[1+ \frac{\exp\left(-\frac{\epsilon^2}{2ng(n)^2}\right)}{g(n)^2}\right]\right) .
\label{eqn:var_pearson_proof_final}
\end{align}

It follows from Corollary 2 in~\cite{li_quantum_2019} that we can approximate $\tilde{E}$ up to an additive error of $\varepsilon/2$ with probability of success of at least $2/3$ using $\widetilde{O}(1/\varepsilon)\cdot 2^{\lceil\log_{2}(\sqrt{n}g(n)/\varepsilon)\rceil} = \widetilde{O}(\sqrt{n}g(n)/\varepsilon^2)$ queries to $q$ and $\widetilde{O}(1/\varepsilon)\cdot 2^{\lceil\log_{2}(\sqrt{n}g(n)^2/\varepsilon)\rceil} = \widetilde{O}(\sqrt{n}g(n)^2/\varepsilon^2)$ queries to $p$. Together with our earlier demonstration that $|E-\tilde{E}|\leq \varepsilon/2$, this completes the proof.
\end{proof}

\end{document}